\documentclass[11pt]{article}
\pdfoutput=1
\usepackage{dcolumn}
\usepackage{bm}

\usepackage[square,numbers,compress,sort]{natbib}
\usepackage{graphicx}
\usepackage{amssymb,amsmath}
\usepackage{multirow}
\usepackage{slashed,xfrac}
\usepackage[makeroom]{cancel}
\usepackage{color,url}
\usepackage[colorlinks=true
,urlcolor=blue
,anchorcolor=blue
,citecolor=blue
,filecolor=blue
,linkcolor=blue
,menucolor=blue
,linktocpage=true
,pdfproducer=medialab
,pdfa=true
]{hyperref}

\usepackage{epsfig,psfrag,rotating,soul}
\usepackage{rotfloat,xcolor}
\usepackage[normalem]{ulem}

\usepackage[textwidth=1.7cm,color=green!85!black,textsize=tiny]{todonotes}

\graphicspath{{figs/}}

\evensidemargin \oddsidemargin
\allowdisplaybreaks

\renewcommand\O{\mathcal O}

\newcommand{\be}{\begin{equation}}
\newcommand{\ee}{\end{equation}}
\newcommand{\ba}{\begin{eqnarray}}
\newcommand{\ea}{\end{eqnarray}}

\def\thefootnote{\fnsymbol{footnote}}

\begin{document}
\thispagestyle{empty}

\begin{flushright}
IFT-UAM/CSIC-18-59\\
FTUAM-18-15\\
LPT-Orsay-18-77
\end{flushright}

\vspace{0.5cm}

\begin{center}

\begin{Large}
\textbf{\textsc{One-loop effective LFV $\boldsymbol{Zl_kl_m}$ vertex from heavy neutrinos within the Mass Insertion Approximation}} 
\end{Large}

\vspace{1cm}

{\sc
 
M.J. Herrero$^1$%
\footnote{\tt \href{mailto:maria.herrero@uam.es}{maria.herrero@uam.es}}%
, X. Marcano$^2$%
\footnote{\tt \href{mailto:xabier.marcano@th.u-psud.fr}{xabier.marcano@th.u-psud.fr}}%
, R. Morales$^3$%
\footnote{\tt \href{mailto:roberto.morales@fisica.unlp.edu.ar}{roberto.morales@fisica.unlp.edu.ar}}%
, A. Szynkman$^3$%
\footnote{\tt \href{mailto:szynkman@fisica.unlp.edu.ar}{szynkman@fisica.unlp.edu.ar}}%
}

\vspace*{.7cm}

{\sl

$^1$Departamento de F\'{\i}sica Te\'orica and Instituto de F\'{\i}sica Te\'orica, IFT-UAM/CSIC,\\
Universidad Aut\'onoma de Madrid, Cantoblanco, 28049 Madrid, Spain

\vspace*{0.1cm}

$^2$Laboratoire de Physique Th\'eorique, CNRS, \\
Univ. Paris-Sud, Universit\'e Paris-Saclay, 91405 Orsay, France

\vspace*{0.1cm}

$^3$IFLP, CONICET - Dpto. de F\'isica, Universidad Nacional de La Plata,\\
C.C. 67, 1900 La Plata, Argentina

}

\end{center}

\vspace*{0.1cm}

\begin{abstract}
\noindent

 In this paper we study the effective lepton flavor violating vertex of an electroweak $Z$ gauge boson and two charged leptons with different flavor, $l_k$ and $l_m$, that is generated to one-loop in low scale seesaw models with right handed neutrinos whose masses are heavier than the electroweak scale. We first compute the form factor describing this vertex by using the mass insertion approximation, where the flavor non-diagonal entries of the neutrino Yukawa   coupling matrix are the unique origin, to one-loop level,  of lepton flavor changing processes with charged leptons in the external legs.  Then, by considering the proper large right handed neutrino mass expansion of the form factor, we derive a formula for the $Z l_k l_m$ effective vertex which is very simple and useful for fast phenomenological estimates. In the last part of this work we focus on the phenomenological applications of this vertex for simple and accurate estimates of the $Z \to l_k {\bar l}_m$ decay rates. Concretely, this vertex will allow us to conclude easily on the maximum allowed decay rates by present data in  the inverse seesaw model. The found rates are promising, at the reach of future lepton colliders. 
 
\end{abstract}

\def\thefootnote{\arabic{footnote}}
\setcounter{page}{0}
\setcounter{footnote}{0}

\newpage
\section{Introduction}
\label{intro}
One of the most interesting aspects of low scale seesaw models \cite{Mohapatra:1986aw,Mohapatra:1986bd,Bernabeu:1987gr,Dittmar:1989yg,GonzalezGarcia:1991be,Pilaftsis:1992st,Ilakovac:1994kj}
with moderately heavy right handed (RH) neutrinos is that they can accommodate easily and successfully the low energy neutrino data, and at the same time they may provide sizable rates for processes with Lepton Flavor Violation (LFV) in the charged lepton sector. The origin of these potential large LFV rates is the allowed large  neutrino Yukawa coupling matrices in these models, $Y_\nu \simeq {\cal O}(1)$, which under the assumption of being non-diagonal in flavor and considering loops involving the RH neutrinos in the internal legs may generate radiatively such LFV processes~\cite{Mann:1983dv,Korner:1992an,Ilakovac:1999md,Illana:1999ww,Illana:2000ic,Arganda:2004bz,Abada:2012cq,Alonso:2012ji,Abada:2012mc,Abada:2013aba,Abada:2014kba,Arganda:2014dta,Abada:2014cca,Abada:2015zea,Arganda:2015naa,Arganda:2015ija,Abada:2015oba,DeRomeri:2016gum,Abada:2016vzu,Arganda:2017vdb}.

Here, we consider the Inverse Seesaw model (ISS) \cite{Mohapatra:1986aw,Mohapatra:1986bd,Bernabeu:1987gr,Dittmar:1989yg}
as a specific realization of these low scale seesaw models, and work with three pairs of RH neutrinos with opposite lepton numbers, which for simplicity are assumed to be quasi-degenerate. The RH mass scale $M_R$ introduced in the ISS by the mass term involving the RH neutrinos is assumed here, also for simplicity, to be flavor diagonal and degenerate in the three diagonal entries. In the present context, we consider this new scale $M_R$ to be above the electroweak (EW) scale, say at the energy interval ${\cal O}(0.1-10)$ TeV, that is accessible at the LHC.   
 
Regarding the specific LFV processes, we focus here on the particular case of the LFV $Z$ boson decays (LFVZD) to charged leptons with different flavor, $Z \to \ell_k {\bar \ell}_m$, which  have also interesting rates in low energy scale seesaw models with heavy neutrinos \cite{Mann:1983dv,Ilakovac:1994kj,Bernabeu:1987gr,Dittmar:1989yg,Korner:1992an,Ilakovac:1999md,Illana:1999ww,Illana:2000ic,Abada:2014cca,DeRomeri:2016gum}, and in particular in the ISS model, as studied in~\cite{Abada:2014cca,DeRomeri:2016gum} working in the physical basis or in~\cite{Abada:2015zea} by computing the relevant Wilson coefficients. 
These decays, as well as the LFV Higgs decays, are being intensely searched for nowadays at the LHC~\cite{Aad:2014bca,Aad:2016blu,Aaboud:2018cxn,Khachatryan:2016rke,Sirunyan:2017xzt} and, the absence of any experimental evidence of these $Z$ decays already sets very stringent bounds on the corresponding decay rates. We summarize in Table \ref{LFVZDexp} the present upper bounds on the various LFVZD channels from both the LEP data and the LHC data. On the other hand, the expectations for improving the sensitivities to these LFVZD rates in the future experiments are quite promising. In particular, the future linear colliders claim an expected sensitivity of $10^{-9}$ \cite{Wilson:I, Wilson:II}, and in the Future Circular  $e^+ e^-$ Colliders (such as FCC-ee (TLEP)\cite{Blondel:2014bra}), where it is estimated that up to $10^{13}$ $Z$ bosons would be produced, the sensitivities could be improved even further.  

The purpose of the present work is to compute the LFVZD rates in the ISS with a different technique: the mass insertion approximation (MIA). The main motivation to use the MIA is that, in contrast to the alternative full one-loop computation \cite{Illana:1999ww,Abada:2014cca,DeRomeri:2016gum}, it provides very simple analytical results and these are written explicitly in terms of the main input parameters of the ISS, concretely, the neutrino right handed mass $M_R$ and the neutrino Yukawa coupling matrix $Y_\nu$. Thus, working directly in the electroweak interaction basis, instead of the physical mass basis, the MIA leads to the simplest results which in turn can be used to further analyze the interesting decoupling behavior of the heavy right handed neutrinos in these LFVZD. We follow here the same method for the MIA as in our previous works \cite{Arganda:2017vdb, Arganda:2015uca} where we applied it to the case of LFV Higgs decays. Other similar methods using the mass insertion technique to compute observables in flavor physics have also been explored in \cite{Dedes:2015twa,Rosiek:2015jua}.  

Our final aim here is to compute the one-loop effective vertex, $Z \ell_k \ell_m$  associated to the proper large $M_R$ expansion of the involved form factors, which will show as a series in powers
of $v^2/M_R^2$, with $v= 174$ GeV characterizing the EW scale. We believe that the simple formulas provided here for this $Z \ell_k \ell_m$ effective vertex can be very useful to test rapidly the compatibility of these  models with LFV data. 
As an illustration of this utility, we will explore here with the obtained  $Z \ell_k \ell_m$ effective vertex the maximum allowed LFVZD rates by present data in a specific low scale seesaw models, the ISS model. We will discuss here that these predicted rates are indeed within the reach of the future experiments.

 The paper is organized as follows: in section~\ref{computationwidth} we summarize the main features of the ISS model in terms of the EW interaction basis, and we present the computation of $\Gamma(Z \to \ell_k \bar{\ell}_m)$ to one-loop within the MIA in all covariant $R_\xi$ gauges and in the unitary gauge. Our proof of the gauge invariance of the on-shell form factor is also included in that section. Section~\ref{computationvertex} contains the computation of the one-loop effective vertex for LFVZD and the comparison of the MIA with the full results.  Section~\ref{maximumLFVZD} is devoted to explore the maximum allowed LFVZD rates using our MIA-effective vertex. The main conclusions are summarized in section~\ref{conclusions}. The technicalities of the present computation, including the conventions for the one-loop integrals, the analytic expressions of the form factors for each diagram of the full and MIA computations in the Feynman-'t Hooft gauge, the expansions of the one-loop functions and the effective vertex at zero external momenta, are collected in the Appendices \ref{LoopIntegrals}, \ref{app:Full}, \ref{app:FormFactors}, \ref{Expansions} and \ref{app:Zpenguins}, respectively.

\begin{table}[t!]
\begin{center}
\begin{tabular}{lllll}
\hline
\hline
LFV Obs. & \multicolumn{4}{c}{Present Upper Bounds  $(95\%~CL)$} \\
\hline
BR$(Z\to\mu e)$ & $1.7\times10^{-6}$ &LEP  (1995)~\cite{Akers:1995gz}\hspace{.7cm} & $7.50\times10^{-7}$ &ATLAS (2014)~\cite{Aad:2014bca}\\
BR$(Z\to\tau e)$ & $9.8\times10^{-6}$ &LEP (1995)~\cite{ Akers:1995gz} & $5.8\times10^{-5}$ &ATLAS (2018)~\cite{Aaboud:2018cxn} \\
BR$(Z\to\tau\mu )$ &$1.2\times10^{-5}$ &LEP (1995)~\cite{Abreu:1996mj}& $1.3\times10^{-5}$ &ATLAS (2018)~\cite{Aaboud:2018cxn}\\
\hline
\hline
\end{tabular}
\caption{Present experimental bounds on Lepton Flavor Violating $Z$ boson decays. 
Here BR$(Z\to \ell_k \ell_m)\equiv \,$BR$(Z\to \ell_k \bar{\ell}_m)+$BR$(Z\to \bar{\ell}_k \ell_m)$.}
\end{center}
\label{LFVZDexp}
\end{table}
%
\section{$\boldsymbol{\Gamma(Z\to\ell_k\bar\ell_m)}$ to one-loop within the MIA}
\label{computationwidth}
Our computation of the partial decay width for the LFVZD in the MIA  is performed in the EW basis. Therefore the starting point is the ISS Lagrangian in the EW basis, i.e.,  in terms of the right and the left handed neutrinos. We follow the same notation and conventions for this Lagrangian as in~\cite{Arganda:2017vdb}: 
\begin{equation}
 \label{ISSlagrangian}
 \mathcal{L}_\mathrm{ISS} = - Y^{ij}_\nu \overline{L_{i}} \widetilde{\Phi} \nu_{Rj} - M_R^{ij} \overline{\nu_{Ri}^c} X_j - \frac{1}{2} \mu_{X}^{ij} \overline{X_{i}^c} X_{j} + h.c.\,,
\end{equation}
where $L$ is the SM lepton doublet, $\widetilde{\Phi}=i\sigma_2\Phi^*$ with $\Phi$ the SM Higgs doublet and 
$i,j$ are indices in flavor space that run from 1 to 3. Correspondingly, $Y_\nu$, $\mu_{X}$ and $M_R$ are $3\times 3$ matrices.
The $C$-conjugate fermion fields are defined here as $f_L^c=(f_L)^c=(f^c)_R$ and $f_R^c=(f_R)^c=(f^c)_L$.

The mass matrix of the ISS, in the EW interaction basis $(\nu_L^c\,,\;\nu_R\,,\;X)$ is: 
\begin{equation}
\label{ISSmatrix}
 M_{\mathrm{ISS}}=\left(\begin{array}{c c c} 0 & m_D & 0 \\ m_D^T & 0 & M_R \\ 0 & M_R^T & \mu_X \end{array}\right)\,,
\end{equation}
with $m_D=v Y_\nu$, and $v = 174\,\mathrm{GeV}$. For simplicity, we choose here $M_R$ diagonal in flavor space and with degenerate diagonal entries\footnote{For a generalization to the non-degenerate case see App.~\ref{app:FormFactors}.}.
 In Appendix \ref{app:Full} we summarize the relevant couplings in the neutrino mass basis. 
 Notice that in the ISS model $\mu_X$ is assumed to be small, related to the smallness of light neutrino masses, and therefore its contributions to our LFV process are negligible~\cite{Arganda:2017vdb}.

 The relation between the neutrino electroweak interaction basis $(\nu_L^c, \nu_R, X)$ and the neutrino mass eigenstate basis, $n_i~(i=1,..,9)$ is given by

\begin{equation}\label{EWtoPhysical}
\left(\begin{array}{c} \nu_L^c \\ \nu_R \\ X \end{array}\right) = U_\nu P_R \left(\begin{array}{c} n_1\\ \vdots \\ n_9\end{array}\right),
\quad
\left(\begin{array}{c} \nu_L \\ \nu_R^c \\ X^c \end{array}\right) = U_\nu^* P_L \left(\begin{array}{c} n_1\\ \vdots \\ n_9\end{array}\right)\,,
\end{equation}
where $U_\nu$ is the rotation matrix leading to the physical neutrino masses $m_{n_i}$, given by:
\begin{equation}
U_\nu^T M_{\rm ISS} U_\nu = {\rm diag}(m_{n_1},\dots,m_{n_9})\,.
\end{equation}
For the charged lepton sector we use the physical mass eigenstate basis in the whole paper.

Next,  we write  the relevant amplitude for these  $Z(p_1) \to \ell_k(-p_2) \bar \ell_m (p_3)$ decays
in terms of the proper form factors with $p_1=p_3-p_2$. In the present case of the ISS with right handed neutrinos  and neglecting the lepton masses there is just one form factor 
involved~\cite{Illana:2000ic}, $F_L$. This will be explicitly shown in our forthcoming computation.  
Thus we write,
\begin{equation}
i {\cal M} = i\epsilon^{\lambda}_{Z}(p_1) \bar{u}_{\ell_k} (-p_2) (F_L \gamma_{\lambda} P_L) v_{\ell_m}(p_3) \,.
\label{ampZtaumu_ISS}
\end{equation}
Then, the partial width is simply  
\be
\Gamma(Z \to \ell_k \bar{\ell}_m) = \frac{m_Z}{24\pi}|F_{L}|^{2}\,,
\ee
where the lepton masses have been safely neglected in the phase space factor.

Now, as explained in~\cite{Arganda:2017vdb}, the MIA computation is organized as an ordered expansion in powers of the neutrino matrix $Y_\nu$.  The non-diagonal elements in flavor space of this matrix, $Y_{\nu}^{ij}$ with $i \neq j$,  are the unique origin of LFV in this ISS context, and each $(vY_{\nu}^{ij})$ factor acts as a mass insertion changing lepton flavor. 
Thus, to a given order in this expansion,  ${\cal O}{(Y_\nu^n)}$,  these off-diagonal entries in the neutrino Yukawa coupling matrix, and via the loops with right handed neutrinos, generate non-vanishing contributions  to the observable of our interest here,  $\Gamma(Z \to \ell_k \bar{\ell}_m)$. 
Specifically, the one-loop form factors receive contributions from leading order (LO) terms of 
$\O(Y_\nu Y_\nu^\dagger)$ ($\O(Y_\nu^2)$ or $\O(Y^2)$, in short) and from next to leading order (NLO) terms of 
 $\O(Y_\nu Y_\nu^\dagger Y_\nu Y_\nu^\dagger)$ ($\O(Y_\nu^4)$ or $\O(Y^4)$, in short).  The systematics to compute these form factors in the MIA is fully explained in our previous work \cite{Arganda:2017vdb} which we follow closely  here. 
 
\begin{figure}[t!]
\begin{center}
\includegraphics[scale=0.9]{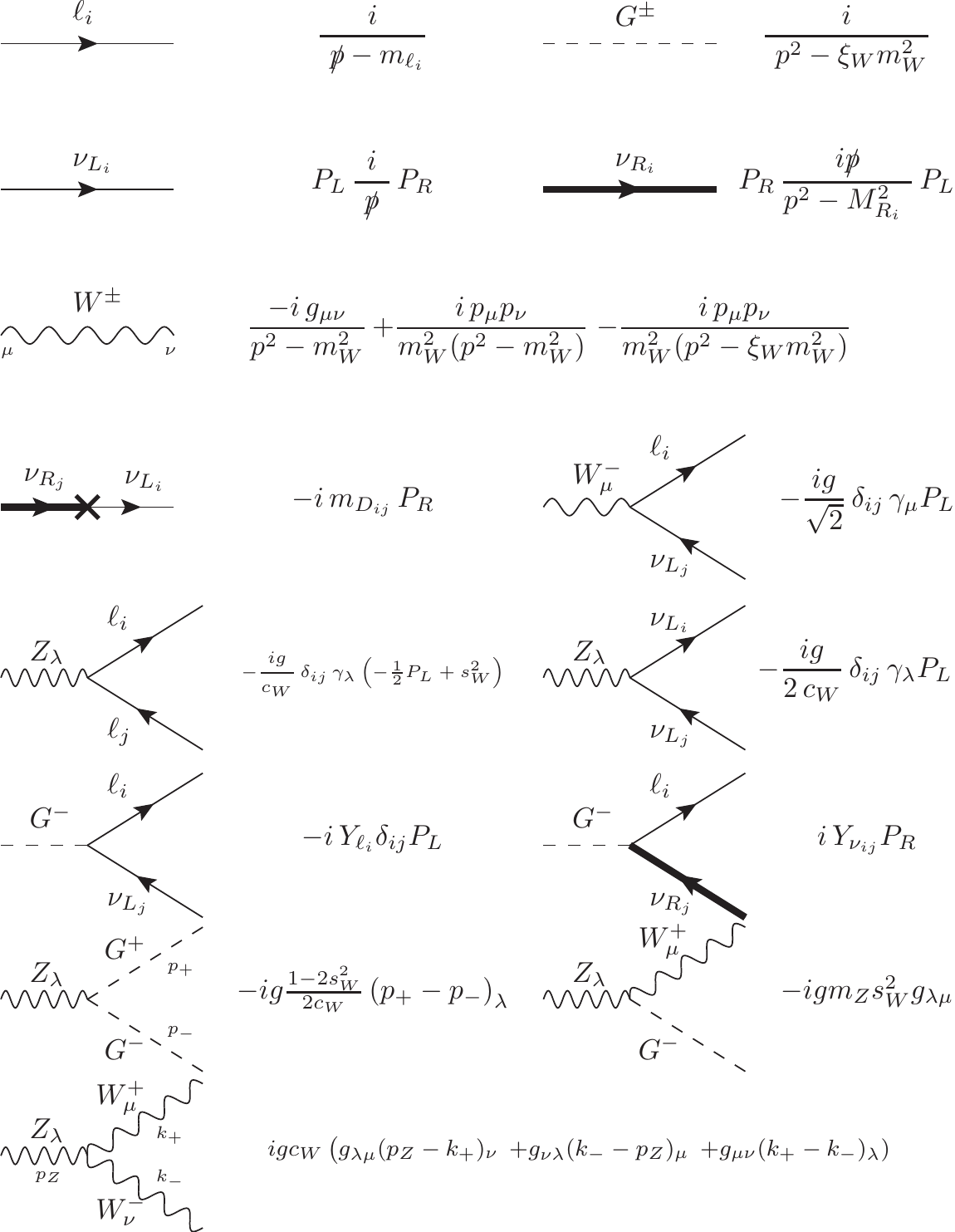}
\caption{Relevant Feynman rules and sign conventions for  the MIA computation in a generic covariant gauge. The momentum convention is that all boson momenta are incoming.  The solid thick line denotes the right handed  neutrino {\it fat} propagator,  as defined and computed in~\cite{Arganda:2017vdb}. The cross denotes our unique LFV insertion given by $m_{D_{ij}}=v Y_{{\nu}_{ij}}$.}  
\label{ZFRall}
\end{center}
\end{figure}

 Our diagrammatic procedure  consists of the systematic insertion of right handed  neutrino ({\it fat}) propagators in all the possible places inside the loops which are built with the relevant interaction vertices and propagators  summarized in Fig.~\ref{ZFRall}. One then follows the counting of the various $Y_\nu$ appearing in each one-loop Feynman diagram. In the present LFVZD these $Y_\nu$  come from just two sources, the coupling of the $\nu_R$  to the Goldstone bosons and the insertions $\nu_L$-$\nu_R$ which go with $m_D=v Y_\nu$.
 Generically, diagrams with one  right handed neutrino propagator will contribute to the form factors at ${\cal O}(Y_{\nu}^{2})$, whereas diagrams with two right handed neutrino propagators will contribute to the form factors at ${\cal O}(Y_{\nu}^{4})$. The detailed computation of the {\it fat} right handed neutrino propagator leading to the expression in Fig.~\ref{ZFRall} can be found in our previous work \cite{Arganda:2017vdb}. The most relevant feature of this {\it fat} propagator is that it contains the resummation of all the insertions given by the  $M_R$ mass insertions (diagonal in flavor) and it neglects the $\mu_X$ mass insertions which are not relevant for the LFV processes of our interest.  
 Thus, dealing with this propagator is very convenient for the present  computation of the LFVZD as it was in \cite{Arganda:2017vdb} for the case the LFV Higgs decays.
It is important to note that there are no couplings between the gauge bosons $W$ and $Z$ with the right handed neutrinos because they are singlets of $SU(2)$. Indeed, these $\nu_R$ can only couple to the Higgs sector, as can be seen in Fig.~\ref{ZFRall}.

\begin{figure}[t!]
\begin{center}
\includegraphics[scale=0.8]{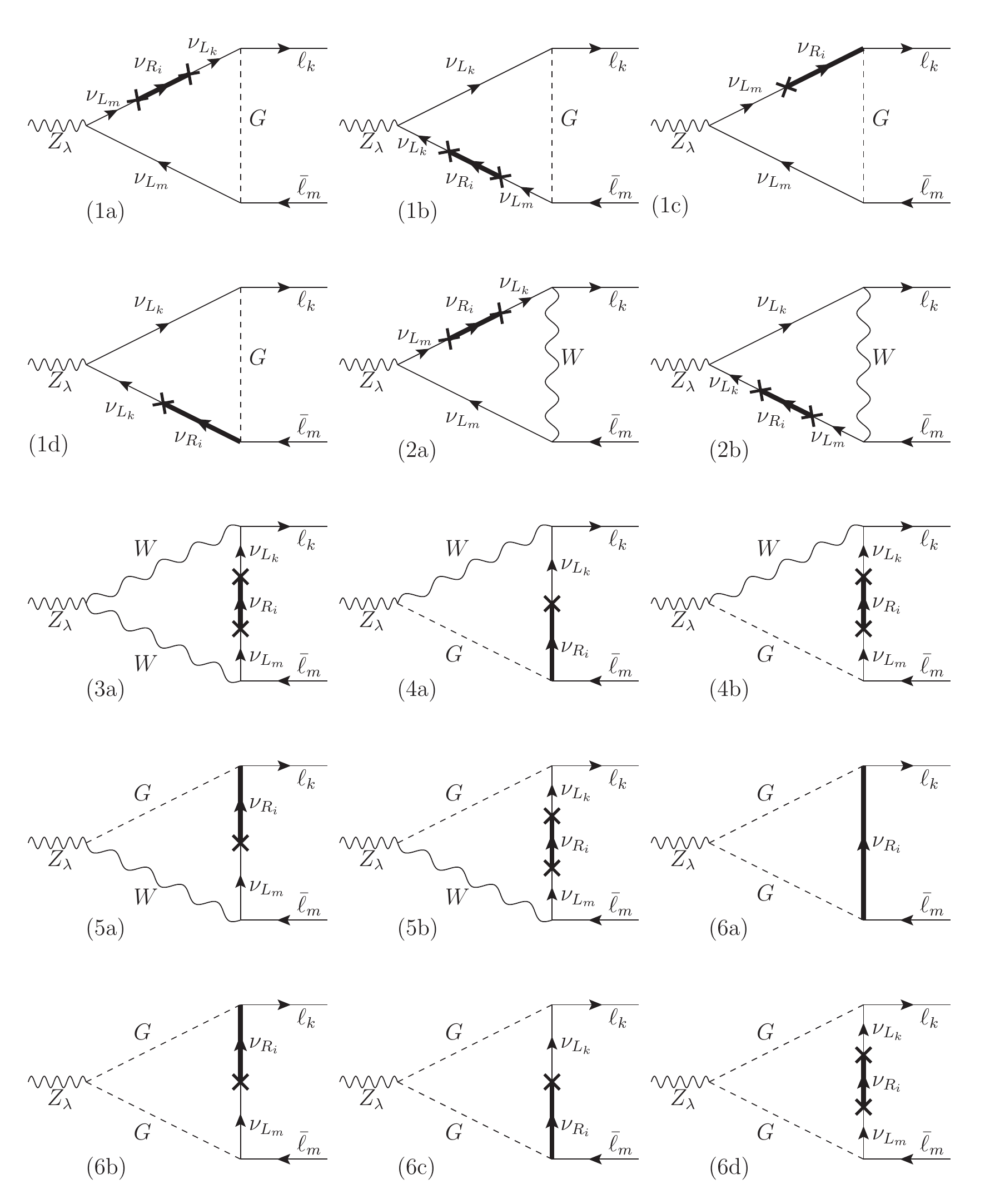}
\caption{Relevant diagrams of $\mathcal O(Y_\nu^2)$ in a covariant gauge corresponding to vertex corrections.}
\label{vertexdiags}
\end{center}
\end{figure}

\begin{figure}[t!]
\begin{center}
\includegraphics[scale=0.8]{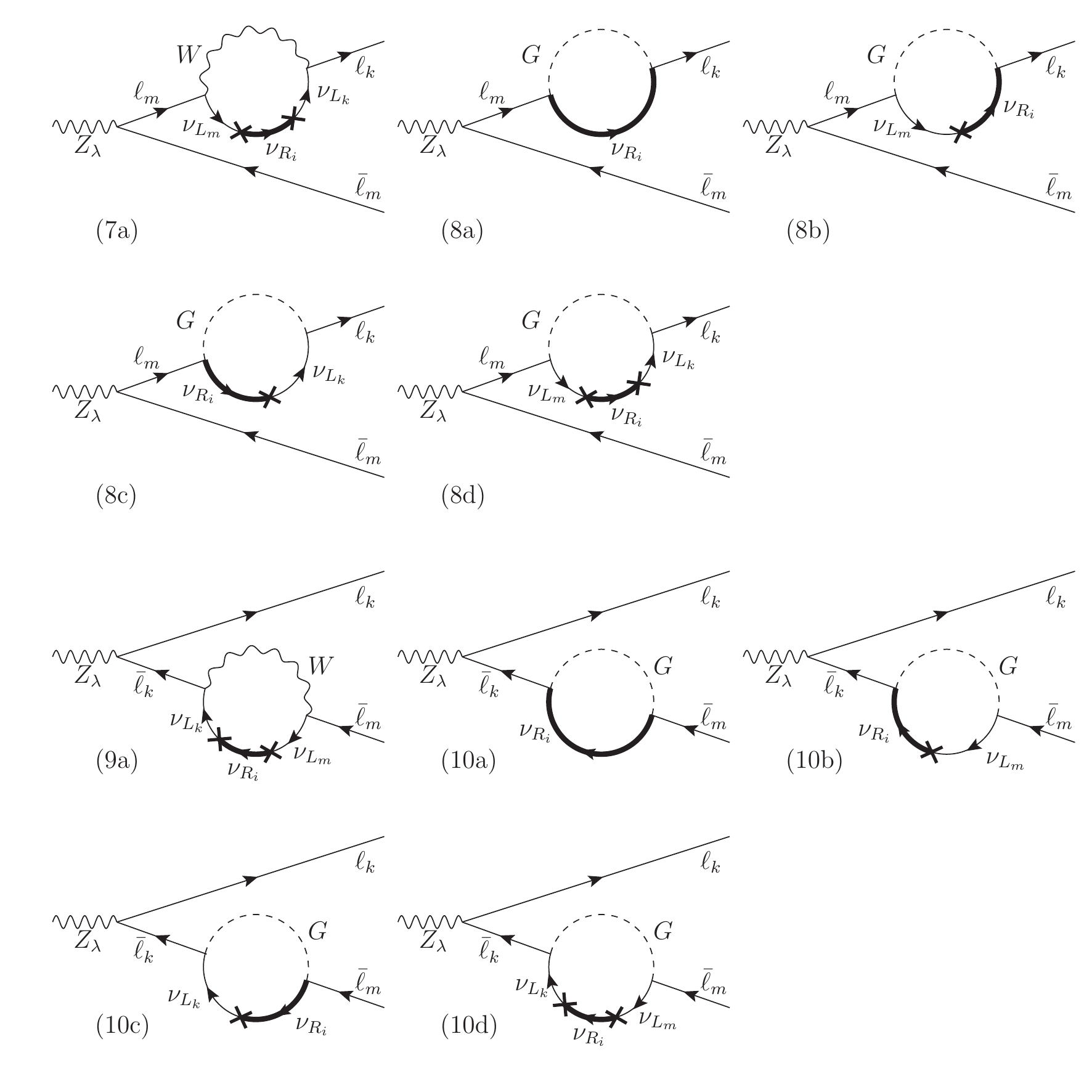}
\caption{Relevant diagrams of $\mathcal O(Y_\nu^2)$ in a covariant gauge corresponding to external leg corrections.}
\label{extlegdiags}
\end{center}
\end{figure}

\begin{figure}[t!]
\begin{center}
\includegraphics[scale=0.8]{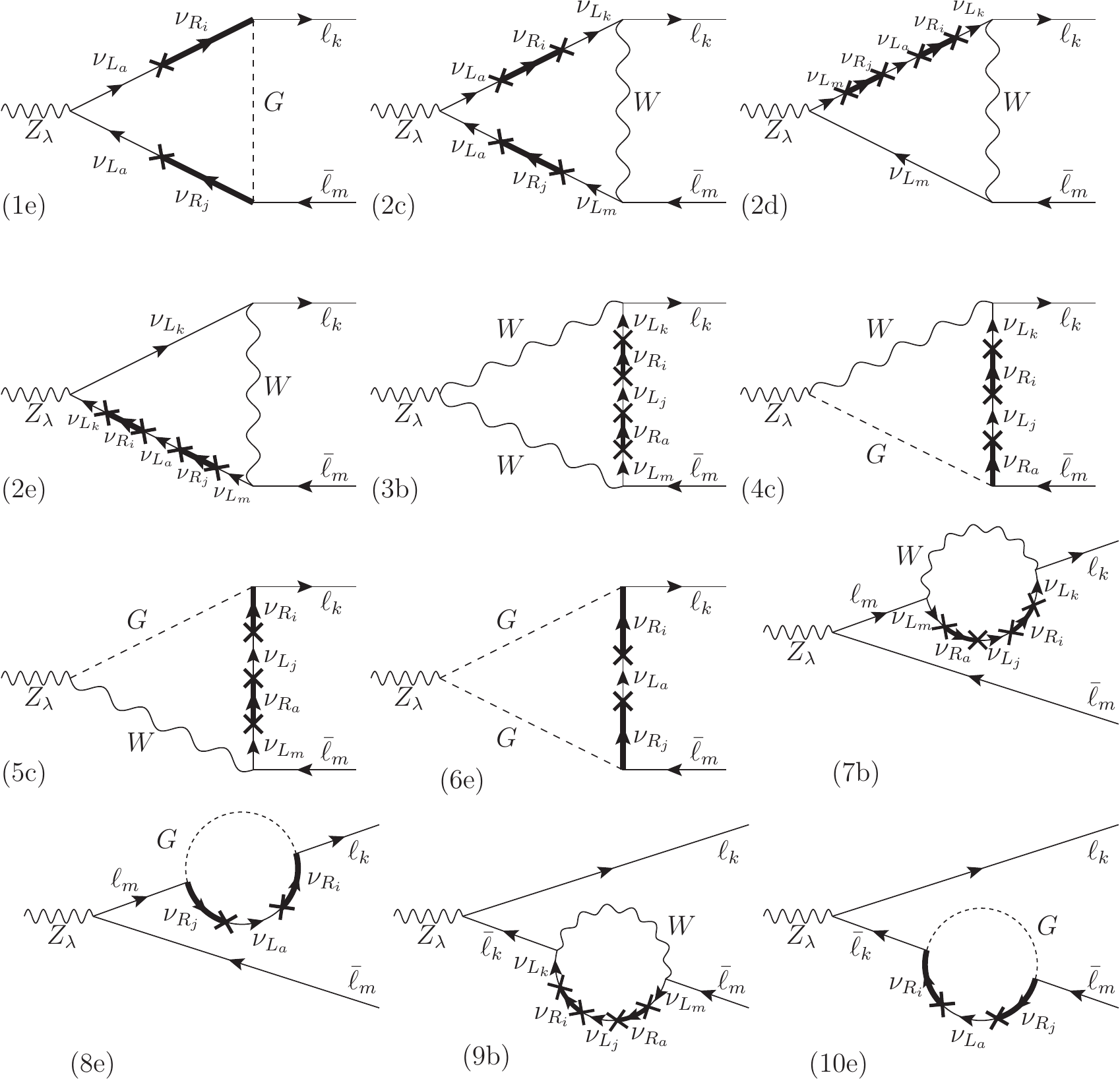}
\caption{Relevant diagrams of $\mathcal O(Y_\nu^4)$ in a covariant gauge.}
\label{dominantY4diags}
\end{center}
\end{figure}

In Figs.~\ref{vertexdiags}, \ref{extlegdiags}, \ref{dominantY4diags}
we show the relevant one-loop diagrams in the MIA corresponding to the dominant contributions of the LO, $\O(Y^2)$,  and the 
NLO, $\O(Y^4)$, respectively, in a generic covariant gauge. Notice that the different topologies in the MIA are of vertex corrections type and of leg corrections type, as in the full computation,  and this suggests our use of a correlated notation for the labelling in the two sets of diagrams, the MIA and the full computation, summarized by diagrams with topology of type (1), type (2), etc.  The final result for the full computation is collected in the Appendix~\ref{app:Full}, for completeness.

The final analytical result for the MIA is the sum of all the contributions  in Figs.~\ref{vertexdiags}, \ref{extlegdiags}, \ref{dominantY4diags}. It gives the total form factor in the MIA to $\O(Y^2+Y^4)$ for arbitrary $Z$ external momentum, $p_1$, that can be summarised as follows:
\ba
F_{L}^{{\rm MIA}}&=&F_{L}^{{\rm MIA\,\, (Y^2)}} +F_{L}^{{\rm MIA\,\, (Y^4)}} \,.
\label{FLtot_FtH}
\ea
At ${\cal O}(Y_{\nu}^{2})$,  the relevant topologies in a covariant gauge are from diagrams all containing 1 right handed neutrino propagator and one of these three combinations: i) 1 vertex with $\nu_R$ and 1 $m_D$ insertion,  
ii)  0 vertices with $\nu_R$ and 2 $m_D$ insertions, iii)  2 vertices with $\nu_R$ and 0 $m_D$ insertions. 
Then,  we get:
\begin{eqnarray}
F_{L}^{{\rm MIA\,\, (Y^2)}} &=&F_{L}^{\rm  (1a)}+F_{L}^{\rm  (1b)}+F_{L}^{\rm  (1c)}+F_{L}^{\rm  (1d)}+F_{L}^{\rm  (2a)}+F_{L}^{\rm  (2b)}+F_{L}^{\rm  (3a)} +F_{L}^{\rm  (4a)}+F_{L}^{\rm  (4b)}
\nonumber \\  
&+& F_{L}^{\rm  (5a)}+F_{L}^{\rm  (5b)}+F_{L}^{\rm  (6a)}+F_{L}^{\rm  (6b)}+F_{L}^{\rm  (6c)}+F_{L}^{\rm  (6d)} 
+F_{L}^{\rm  (7a)}+ F_{L}^{\rm  (8a)}+F_{L}^{\rm  (8b)}
\nonumber \\ 
&+&F_{L}^{\rm  (8c)}+F_{L}^{\rm  (8d)} + F_{L}^{\rm  (9a)}+ F_{L}^{\rm  (10a)}+F_{L}^{\rm  (10b)}+F_{L}^{\rm  (10c)}+F_{L}^{\rm  (10d)} . 
\label{FFLMIAY2}
\end{eqnarray}
The explicit analytical results in the Feynman-'t Hooft gauge for all the relevant  diagrams are collected in Appendix \ref{app:FormFactors}. Notice that some diagrams are subleading since they are of  $\O(m_{\rm lep}^2)$ and their contributions will be neglected from now on.

At ${\cal O}(Y_{\nu}^{4})$,  the relevant topologies in a covariant gauge are from diagrams all containing 2 right handed neutrino propagators and one of these three combinations: i) 2 vertices with $\nu_R$ and 2 $m_D$ insertions,  
ii)  0 vertices with $\nu_R$ and 4 $m_D$ insertions, iii)  1 vertex with $\nu_R$ and 3 $m_D$ insertions. 
Thus, we find that in the Feynman-'t Hooft gauge the most relevant diagrams are those of type (1), (6), (8) and (10) summarized in Fig.~\ref{dominantY4diags} whose respective contributions are given by:
\begin{eqnarray}
F_{L}^{{\rm MIA \,\, (Y^4)}}&=&F_{L}^{\rm  (1e)}+F_{L}^{\rm  (6e)}+F_{L}^{\rm  (8e)}+F_{L}^{\rm  (10e)}\,,
 \label{FFLMIAY4}
\end{eqnarray}
 and the explicit analytical results in the Feynman-'t Hooft gauge for all the relevant diagrams are collected in Appendix \ref{app:FormFactors}.
 The rest of diagrams in Fig.~\ref{dominantY4diags} do not provide leading contributions in the large $M_R$ expansion in the Feynman-'t Hooft gauge, but they will contribute in a generic covariant gauge, as it will be shown in the next subsection. 
 Specifically, the diagrams (2c), (2d), (2e), (3b), (4c), (5c), (7b) and (9b), after performing the expansion in inverse powers of $M_R$ give contributions to the form factor in the Feynman-'t Hooft gauge of ${ \cal O}(v^4/M_R^4)$, whereas they give contributions of ${ \cal O}(v^2/M_R^2)$
in a generic covariant gauge.
 Besides, diagrams with more than two right handed neutrino propagators will also provide subleading corrections in the heavy $M_R$ case of our interest, since they will come with extra powers of $M_R$ in the denominator. Then we neglect contributions of ${\cal O}(Y_{\nu}^{6})$ and higher terms.

All the previous results are given in terms of the standard loop functions whose definitions and conventions are collected in Appendix \ref{LoopIntegrals}. The discussion of the results for other gauge choices like the unitary gauge and the general $R_\xi$ covariant gauges is included in the next subsection \ref{gaugeinvariance}, where our proof of gauge invariance of the on-shell $Z$ form factor for LFVZD will be presented.

The final analytical result is the sum of all these contributions in Eqs.~\eqref{FFLMIAY2} and \eqref{FFLMIAY4}, and this gives the total form factor in the MIA  for arbitrary $Z$ external momentum, $p_1$, in the Feynman-'t Hooft gauge. Then, neglecting the fermion masses $m_k$ and $m_m$, we finally get:
\begin{align}
F_{L}^{{\rm MIA\,\, (Y^2)}} 
&= \frac{1}{16 \pi^{2}} \frac{g}{c_W} \left(Y^{}_{\nu} Y_{\nu}^{\dagger} \right)^{km} \bigg\{ \left(\frac{1}{2}-s_{W}^{2}\right)B_1(M_R,m_W) +(1-2s_{W}^{2})C_{00}(p_2,p_1,M_R,m_W,m_W)   \nonumber\\
&  +2m_{W}^{2}C_{0}(p_2,p_1,M_R,m_W,m_W) +\left(1-2s_{W}^{2}\right)m_{W}^{2}C_{2}(0,M_R,m_W)    \nonumber\\
&  +m_{W}^{2}\big( -2D_{00}+p_{1}^{2}(D_0+D_1+D_{13}-D_{33})\big)(p_{2},0,p_{1},m_{W},0,M_{R},0)   \nonumber\\
&  +m_{W}^{2}\big( -2D_{00}+p_{1}^{2}(D_0+D_1+D_{12}-D_{22})\big)(p_{2},p_{1},0,m_{W},0,M_{R},0)    \nonumber\\
&  +2c_{W}^{2}m_{W}^{2}(2D_{00}-p_{1}^{2}D_{2})(0,p_{2},p_{1},0,M_{R},m_{W},m_{W})  \bigg\}\,,
\label{FLtot_FtH_Y2}
\end{align}
and:
\begin{align}
F_{L}^{{\rm MIA\,\, (Y^4)}} 
  &= \frac{1}{16 \pi^{2}} \frac{g}{c_W} v^{2} \left(Y^{}_{\nu} Y_{\nu}^{\dagger} Y^{}_{\nu} Y_{\nu}^{\dagger} \right)^{km} \bigg\{ \frac{1}{2}C_{0}(p_{2},p_{1},m_W,M_{R},M_R)   \nonumber\\
&  + \left( \frac{1}{2}-s_{W}^{2} \right) \big( 2D_{00}(0,p_{2},p_{1},M_R,M_{R},m_W,m_W) + C_2(M_R,M_{R},m_W) \big) \bigg\}\,.
\label{FLtot_FtH_Y4}
\end{align}
We have also checked that the total MIA form factor in the Feynman-'t Hooft gauge, $F_L^{\rm MIA}$, presented in Eqs.~\eqref{FLtot_FtH}, \eqref{FLtot_FtH_Y2} and \eqref{FLtot_FtH_Y4}, is finite. In fact, the only divergent diagrams in the MIA computation are (6a), (8a) and (10a), and we have proven that the divergences cancel out when adding these three diagrams.
  
We are mainly interested here in the form factor for the LFVZD, i.e.  when the $Z$ boson is on-shell. Then, the proper form factor is obtained by setting $p_1^2=m_Z^2$ in the previous equations. We get:
\begin{align}
F_{L}^{{\rm MIA\,\, (Y^2)}}|_{p_1^2=m_Z^2}  
&= \frac{1}{16 \pi^{2}} \frac{g}{c_W} \left(Y^{}_{\nu} Y_{\nu}^{\dagger} \right)^{km} 
\bigg\{ \left(\frac{1}{2}-s_{W}^{2}\right)B_1(M_R,m_W)   \nonumber\\
& +(1-2s_{W}^{2}){\hat C}_{00}(p_2,p_1,M_R,m_W,m_W)   \nonumber\\
& \left. +2m_{W}^{2}{\hat C}_{0}(p_2,p_1,M_R,m_W,m_W) +\left(1-2s_{W}^{2}\right)m_{W}^{2}C_{2}(0,M_R,m_W)   \right. \nonumber\\
& \left. +m_{W}^{2}\big( -2{\hat D}_{00}+m_{Z}^{2}({\hat D}_0+{\hat D}_1+{\hat D}_{13}-{\hat D}_{33})\big)(p_{2},0,p_{1},m_{W},0,M_{R},0)  \right. \nonumber\\
& \left. +m_{W}^{2}\big( -2{\hat D}_{00}+m_{Z}^{2}({\hat D}_0+{\hat D}_1+{\hat D}_{12}-{\hat D}_{22})\big)(p_{2},p_{1},0,m_{W},0,M_{R},0)   \right. \nonumber\\
&  +2c_{W}^{2}m_{W}^{2}(2{\hat D}_{00}-m_{Z}^{2}{\hat D}_{2})(0,p_{2},p_{1},0,M_{R},m_{W},m_{W})  \bigg\}\,,
\label{FLtot_Zonshell_FtH_Y2}
\end{align}
and:
\begin{align}
F_{L}^{{\rm MIA\,\, (Y^4)}}|_{p_1^2=m_Z^2} 
  &= \frac{1}{16 \pi^{2}} \frac{g}{c_W} v^{2} \left(Y^{}_{\nu} Y_{\nu}^{\dagger} Y^{}_{\nu} Y_{\nu}^{\dagger} \right)^{km} \bigg\{ \frac{1}{2}
  {\hat C}_{0}(p_{2},p_{1},m_W,M_{R},M_R)   \nonumber\\
&  + \left( \frac{1}{2}-s_{W}^{2} \right) \left( 2{\hat D}_{00}(0,p_{2},p_{1},M_R,M_{R},m_W,m_W) + C_2(M_R,M_{R},m_W) \right) \bigg\}\,,
\label{FLtot_Zonshell_FtH_Y4}
\end{align}
where the loop functions with a `hat' means that they are evaluated at $p_1^2=m_Z^2$ (and, $p_2^2= p_3^2 = 0$, since we are neglecting the lepton masses). The definitions of all these loop functions above and their interesting limits for the present paper are collected in Appendices \ref{LoopIntegrals} and \ref{Expansions}.
\subsection{Check of gauge invariance of the on-shell form factor for LFVZD}
\label{gaugeinvariance}

An interesting check of our results in Eqs.~\eqref{FLtot_Zonshell_FtH_Y2} and \eqref{FLtot_Zonshell_FtH_Y4} for the form factor $F_L$ of the $Z$ decay in the Feynman-'t Hooft gauge is to verify the equivalence with the results in the unitary gauge (UG) and with an arbitrary covariant $R_\xi$ gauge.
In order to perform a systematic computation, we start with the UG's calculation.
We consider the contributions of $\mathcal O(Y_\nu^2)$ and $\mathcal O(Y_\nu^4)$ corresponding only to diagrams of type (2), (3), (7) and (9) of Figs.~\ref{vertexdiags}, \ref{extlegdiags} and \ref{dominantY4diags} because there are not Goldstone bosons in this gauge.
We then split the propagator $P_W^{\rm UG}$ of the $W$ gauge boson into two parts, $P_W^{a}$ and $P_W^{b}$:
\be
P_W^{\rm UG}=P_W^{a}+P_W^{b}=-\frac{i g_{\mu \nu}}{p^{2}-m_{W}^{2}} +\frac{i p_{\mu}p_{\nu}}{m_{W}^{2}(p^{2}-m_{W}^{2})} \, .
\ee
In this way, the first part $P_W^{a}$ matches with the propagator of the $W$ gauge boson in the Feynman-'t Hooft gauge. 
Thus, we classify the different contributions arising from each dominant diagram. 
Those contributions at $\mathcal O(Y_\nu^2)$ coming from $P_W^{a}$ are the same contributions as in the Feynman-'t Hooft gauge.
In addition to them, there are new contributions corresponding to $P_W^{b}$ in diagrams (2a), (2b), (3a), (7a) and (9a) of Figs.~\ref{vertexdiags} and \ref{extlegdiags}.
On the other hand, the dominant contributions of $\mathcal O(Y_\nu^4)$ correspond to $P_W^{b}$ in diagrams (2c), (2d), (2e), (3b), (7b) and (9b) of Fig.~\ref{dominantY4diags}.
In the following, we show the result of the form factor $F_L$ in the UG at $\mathcal O(Y_\nu^2+Y_\nu^4)$ for arbitrary $Z$ external momentum $p_1$:
\be
F_{L}^{{\rm UG}}=F_{L}^{{\rm UG\,\, (Y^2)}}+F_{L}^{{\rm UG\,\, (Y^4)}},
\label{FLUG}
\ee
where,
\begin{align}
F_{L}^{{\rm UG\,\, (Y^2)}} &= \frac{1}{16 \pi^{2}} \frac{g}{c_W} \left(Y^{}_{\nu} Y_{\nu}^{\dagger} \right)^{km} \bigg\{ \left(\frac{1}{2}-s_{W}^{2}\right)B_1(M_R,m_W)   \nonumber\\
&+\left(2c_{W}^{2}-\frac{p_{1}^{2}}{m_{Z}^{2}}\right)C_{00}(p_2,p_1,M_R,m_W,m_W)   \nonumber\\
& \left. +2c_{W}^{2}p_{1}^{2}C_{0}(p_2,p_1,M_R,m_W,m_W) +\left(1-2s_{W}^{2}\right)m_{W}^{2}C_{2}(0,M_R,m_W)   \right. \nonumber\\
& \left. +m_{W}^{2}( -2D_{00}+p_{1}^{2}(D_0+D_1+D_{13}-D_{33}))(p_{2},0,p_{1},m_{W},0,M_{R},0)  \right. \nonumber\\
& \left. +m_{W}^{2}( -2D_{00}+p_{1}^{2}(D_0+D_1+D_{12}-D_{22}))(p_{2},p_{1},0,m_{W},0,M_{R},0)   \right. \nonumber\\
&  +2c_{W}^{2}m_{W}^{2}(2D_{00}-p_{1}^{2}D_{2})(0,p_{2},p_{1},0,M_{R},m_{W},m_{W})  \bigg\} \, ,
\label{FLtot_U_Y2}
\end{align}
and,
\begin{align}
F_{L}^{{\rm UG\,\, (Y^4)}} &= \frac{1}{16 \pi^{2}} \frac{g}{c_W} v^{2} \left(Y^{}_{\nu} Y_{\nu}^{\dagger} Y^{}_{\nu} Y_{\nu}^{\dagger} \right)^{km} \bigg\{ \frac{1}{2}C_{0}(p_{2},p_{1},m_W,M_{R},M_R)   \nonumber\\
& \left. + \left( 2c_{W}^{2}-\frac{p_1^{2}}{m_Z^{2}} \right) D_{00}(0,p_{2},p_{1},M_R,M_{R},m_W,m_W) \right. \nonumber\\
&  + \left( \frac{1}{2}-s_{W}^{2} \right) C_2(M_R,M_{R},m_W) \bigg\} \, .
\label{FLtot_U_Y4}
\end{align}
 
These general expressions are not equal to Eqs.~\eqref{FLtot_FtH_Y2} and \eqref{FLtot_FtH_Y4}, but for the particular case of the $Z$ decays (setting $p_1^2=m_Z^2$ in the previous equations) and using the relations $s_W^2+c_W^2=1$ and $m_W^2=c_W^2m_Z^2$, we arrive to Eqs.~\eqref{FLtot_Zonshell_FtH_Y2} and \eqref{FLtot_Zonshell_FtH_Y4} exactly.
Thus, we verify the equivalence of the form factors for the $Z$ decay in the Feynman-'t Hooft gauge and the UG when the $Z$ is on-shell.
\newline

Finally, we present the computation in an arbitrary covariant $R_\xi$ gauge.
 The relevant gauge-fixing parameter $\xi \equiv \xi_W$ corresponds to $SU(2)_L$ gauge group.
In this case, we may split the propagator into three parts, the two previous ones, $P_W^{a}$ and $P_W^{b}$, and a new part, $P_W^{c}$, that contains the $\xi$ dependence:  
\be
P_W^{\rm R_\xi}=P_W^{a}+P_W^{b}+P_W^{c}=-\frac{i g_{\mu \nu}}{p^{2}-m_{W}^{2}} +\frac{i p_{\mu}p_{\nu}}{m_{W}^{2}(p^{2}-m_{W}^{2})} -\frac{i p_{\mu}p_{\nu}}{m_{W}^{2}(p^{2}-\xi m_{W}^{2})} \, .
\ee
The contribution to the form factor from $P_W^{c}$ can be easily obtained from that of $P_W^{b}$ by simply changing the global sign and by replacing $m_W \to \sqrt{\xi} m_W$ in the corresponding argument of masses of the involved one-loop functions.
 
On the other hand, the Goldstone boson propagator is now given by:
\be
P_G^{\rm R_\xi}=\frac{i}{p^{2}-\xi m_{W}^{2}} \, ,
\ee
thus, in comparison with the previous computation of the Feynman-'t Hooft gauge, we now 
have to replace $m_W \to \sqrt{\xi} m_W$ in the masses of the one-loop functions corresponding to the Goldstone bosons.

Taking into account all the above commented  properties, we find the form factor $F_L$ in an arbitrary covariant $R_\xi$ gauge at $\mathcal O(Y_\nu^2+Y_\nu^4)$, for arbitrary $Z$ external momentum $p_1$, from the computation of all diagrams in Figs.~\ref{vertexdiags}, \ref{extlegdiags}, \ref{dominantY4diags}. This is given by:
\be
F_{L}^{{\rm R_\xi}}=F_{L}^{{\rm R_\xi\,\, (Y^2)}}+F_{L}^{{\rm R_\xi\,\, (Y^4)}},
\label{FLtot_xi}
\ee
where,
\begin{align}
F_{L}^{{\rm R_\xi\,\, (Y^2)}} &= \frac{1}{16 \pi^{2}} \frac{g}{c_W} \left(Y^{}_{\nu} Y_{\nu}^{\dagger} \right)^{km} \bigg\{ \left(2c_{W}^{2}B_0+\left(\frac{1}{2}-s_{W}^{2}\right)B_1\right)(M_R,m_W)    \nonumber\\
& \left. -\left(\frac{p_{1}^{2}}{m_{Z}^{2}}-2c_{W}^{2}\right)C_{00}(p_2,p_1,M_R,m_W,m_W)   \right. \nonumber\\
& \left. +2c_{W}^{2}p_{1}^{2}C_{0}(p_2,p_1,M_R,m_W,m_W) +\left(1-2s_{W}^{2}\right)m_{W}^{2}C_{2}(0,M_R,m_W)   \right. \nonumber\\
& \left. +m_{W}^{2}( -2D_{00}+p_{1}^{2}(D_0+D_1+D_{13}-D_{33}))(p_{2},0,p_{1},m_{W},0,M_{R},0)  \right. \nonumber\\
& \left. +m_{W}^{2}( -2D_{00}+p_{1}^{2}(D_0+D_1+D_{12}-D_{22}))(p_{2},p_{1},0,m_{W},0,M_{R},0)   \right. \nonumber\\
& \left. +2c_{W}^{2}m_{W}^{2}(2D_{00}-p_{1}^{2}D_{2})(0,p_{2},p_{1},0,M_{R},m_{W},m_{W})  \right. \nonumber\\
& \left. -2c_{W}^{2}B_0(M_R,\sqrt{\xi}m_W) +\left(1-\frac{p_{1}^{2}}{m_{Z}^{2}}\right)C_{00}(p_2,p_1,M_R,\sqrt{\xi}m_W,\sqrt{\xi}m_W)  \right. \nonumber\\
& \left. +\left(\frac{p_{1}^{2}}{m_{Z}^{2}}-1\right)C_{00}(p_2,p_1,M_R,\sqrt{\xi}m_W,m_W)  +\left(\frac{p_{1}^{2}}{m_{Z}^{2}}-1\right)C_{00}(p_2,p_1,M_R,m_W,\sqrt{\xi}m_W)  \right. \nonumber\\
& \left. +\left(\xi c_W^{2}m_W^{2}-c_W^{2}p_1^2+s_W^{2}m_W^{2}\right)C_0(p_2,p_1,M_R,\sqrt{\xi}m_W,m_W)  \right. \nonumber\\
&  +\left(\xi c_W^{2}m_W^{2}-c_W^{2}p_1^2+s_W^{2}m_W^{2}\right)C_0(p_2,p_1,M_R,m_W,\sqrt{\xi}m_W)  \bigg\} \, ,
\label{FLtotxiY2}
\end{align}
and,
\begin{align}
F_{L}^{{\rm R_\xi\,\, (Y^4)}} &= \frac{1}{16 \pi^{2}} \frac{g}{c_W} v^{2} \left(Y^{}_{\nu} Y_{\nu}^{\dagger} Y^{}_{\nu} Y_{\nu}^{\dagger} \right)^{km} \bigg\{ \frac{1}{2}C_{0}(p_{2},p_{1},m_W,M_{R},M_R) +2c_W^{2}C_{0}(M_{R},M_R,m_W)   \nonumber\\
&\left. -\left( \frac{p_1^2}{m_Z^{2}}-2c_W^{2} \right)D_{00}(0,p_{2},p_{1},M_R,M_{R},m_W,m_W) +\left( \frac{1}{2}-s_{W}^{2} \right)C_2(M_R,M_{R},m_W) \right.  \nonumber\\
&\left. -2c_W^{2}C_0(M_R,M_{R},\sqrt{\xi}m_W) +\left(1-\frac{p_1^2}{m_Z^{2}}\right)D_{00}(0,p_{2},p_{1},M_R,M_{R},\sqrt{\xi}m_W,\sqrt{\xi}m_W) \right.  \nonumber\\
& \left. +\left( \left( \frac{p_1^2}{m_Z^{2}}-1 \right)D_{00} +\left( \xi c_W^{2}m_W^{2}-c_W^{2}p_1^2+s_W^{2}m_W^{2} \right)D_0 \right)(0,p_{2},p_{1},M_R,M_{R},\sqrt{\xi}m_W,m_W) \right.  \nonumber\\  
&  +\left( \left( \frac{p_1^2}{m_Z^{2}}-1 \right)D_{00} +\left( \xi c_W^{2}m_W^{2}-c_W^{2}p_1^2+s_W^{2}m_W^{2} \right)D_0 \right)(0,p_{2},p_{1},M_R,M_{R},m_W,\sqrt{\xi}m_W) \bigg\} \,. 
\label{FLtotxiY4}
\end{align}

On the other hand, from the integral definition of the one-loop functions (see Appendix \ref{LoopIntegrals}), we obtain the following useful relations: 
\ba
&2c_{W}^{2}\left(B_0(M_R,m_W)-B_0(M_R,\sqrt{\xi}m_W)\right)-\left(1 -\xi \right)c_{W}^{2}m_{W}^{2}C_0(p_2,p_1,M_R,\sqrt{\xi}m_W,m_W)  \nonumber\\
&-\left(1 -\xi \right)c_{W}^{2}m_{W}^{2}C_0(p_2,p_1,M_R,m_W,\sqrt{\xi}m_W) \approx \mathcal O(m_{lep}^{2}) \approx 0 \, ,
\label{invgaugeY2}
\ea

and 
 \ba
&2c_{W}^{2}\left(C_0(M_R,M_R,m_W)-C_0(M_R,M_R,\sqrt{\xi}m_W)\right)  \nonumber\\
&-\left(1 -\xi \right)c_{W}^{2}m_{W}^{2}D_0(0,p_2,p_1,M_R,M_R,\sqrt{\xi}m_W,m_W)  \nonumber\\
&-\left(1 -\xi \right)c_{W}^{2}m_{W}^{2}D_0(0,p_2,p_1,M_R,M_R,m_W,\sqrt{\xi}m_W) \approx \mathcal O(m_{lep}^{2}) \approx 0 \, ,
\label{invgaugeY4}
\ea
which can be used to further simplify the previous results of $\mathcal O(Y_\nu^2)$ and $\mathcal O(Y_\nu^4)$ respectively.
It is worth noticing that the general result for an arbitrary $p_1^2$ is not gauge invariant, since the $\xi$ dependence is not fully cancelled in Eqs.~(\ref{FLtotxiY2}) and (\ref{FLtotxiY4}).
 
Finally, to get the form factors in an arbitrary $R_\xi$ gauge in the case where the $Z$ gauge boson is on-shell, we set $p_1^2= m_Z^ 2$ in Eqs.~\eqref{FLtotxiY2} and \eqref{FLtotxiY4}, and  use the above relations in Eqs.~\eqref{invgaugeY2} and \eqref{invgaugeY4}. Simplifying the final expression by means of  the identities $s_W^2+c_W^2=1$ and $m_W^2=c_W^2m_Z^2$, we find out that the expected cancellations among the $\xi$ dependent terms take place and the final result for  
$F_{L}^{{\rm R_\xi }}|_{p_1^2=m_Z^2}$ turns out to be $\xi$ independent, leading to the same result for all $\xi$ choices and coinciding with the results of the previous section for the Feynman-'t Hooft gauge, 
$F_{L}^{{\rm MIA\,\, (Y^2)}}|_{p_1^2=m_Z^2}$ and $F_{L}^{{\rm MIA\,\, (Y^4)}}|_{p_1^2=m_Z^2}$ in Eqs.~\eqref{FLtot_Zonshell_FtH_Y2} and \ref{FLtot_Zonshell_FtH_Y4}, respectively.
Therefore, having found the same result for all $R_\xi$ gauges as well as for the unitary gauge, we conclude that our result for the on-shell form factor describing the LFVZD, 
$F_L|_{p_1^2=m_Z^2}$, is gauge invariant. This is as expected, since this is a physical quantity defining an observable, the partial LFVZD width.  

Another interesting discussion arises when considering the zero external momenta approximation, and testing if this approximation is or is not appropriate to estimate the LFVZD rates. We have also explored  this question in detail in this work, and our conclusion is that it is not appropriate, because it is not a gauge invariant quantity. Since this issue is not needed for the central results in this paper, we present this discussion separately, and leave it to Appendix \ref{app:Zpenguins}. There we present the results of the form factors in the case of zero $Z$ external momentum, $p_1^2 = 0$, and include our proof that the result in that case is not gauge invariant. This is in contrast, with the on-shell case presented in this section, being fully gauge invariant. 

 \section{Computation of the one-loop effective vertex for LFVZD}
\label{computationvertex}
Here we present the computation of the one-loop effective vertex that is the proper one for the description of  the LFVZD. This leads us to set first the $Z$ external leg to be on-shell, and second to explore the proper analytic expansion of the MIA form factor $F_L^{\rm MIA}$ that is valid at large $M_R$. The effective vertex that we look for then summarizes the one-loop effects of the heavy right handed neutrinos, and it is obtained from the result of this large $M_R$ expansion, generically, as:
\be 
{\hat V}^{\rm eff}_{Z\ell_k\ell_m}=V^{\rm eff}_{Z\ell_k\ell_m}|_{p_1^2=m_Z^2} 
= F_L^{\rm MIA}|_{p_1^2=m_Z^2} ({\bf M_R}),
\ee 
where the mass ${\bf M_R}$ in boldface means that the function 
$F_L^{\rm MIA}|_{p_1^2=m_Z^2}$ has been expanded at large $M_R$. Specifically, this expansion should valid for $M_R\gg v$, with $v=174$ GeV being the characteristic scale providing all the electroweak masses involved. In practice, this implies $M_R$ being much heavier than all the other particle masses involved in the loop contributing diagrams, therefore, larger 
than $m_Z$, $m_W$, $m_{\rm lep}$, etc. 

The result of this large $M_R$ expansion
must provide a local function in spacetime, hence, leading in momentum space, to an expansion (up to logarithms) given in inverse powers of $M_R^2$. 
In fact, our explicit computation presented in this work shows that the first term in this expansion is of $\O(v^2/M_R^2)$, the second term is of $\O(v^4/M_R^4)$ and so on, leading to an expansion, valid at large $M_R \gg v$,  of the generic form:
\be 
{\hat V}^{\rm eff}_{Z\ell_k\ell_m} \simeq c_1 \frac{v^2}{M_R^2} + c_2 \frac{v^4}{M_R^4}+...,
\label{generic}
\ee
where the coefficients $c_{1,2,\dots}$ contain contributions that are generically either constants with $M_R$ or logarithmic contributions like $\log(m_W^2/M_R^2)$, which are originated from the heavy neutrino loops with $W$ gauge bosons  and/or Goldstone bosons.   
This functional dependence with $M_R$ typically signals the decoupling behavior of the heavy neutrinos in the one-loop generated LFV form factors, leading to one-loop radiative effects in the LFVZD decays that vanish in the asymptotically infinite right handed mass limit. 

In order to get this expansion of the on-shell form factors in inverse powers of $M_R^2$, we start with the previous results of the Feynman-'t Hooft gauge, $F_{L}^{{\rm MIA\,\, (Y^2+Y^4)}}$, in Eqs.~\eqref{FLtot_Zonshell_FtH_Y2} and \eqref{FLtot_Zonshell_FtH_Y4} and insert into these equations the corresponding expansions for the loop functions that we have also computed and whose results are collected in Appendix \ref{Expansions}. 
This leads us to very  simple results for the on-shell form factors, which in turn define the wanted on-shell effective vertex, such that:  

\begin{equation}
i {\cal M} \simeq  i \epsilon^{\lambda}_{Z}(p_1) \bar{u}_{\ell_k} (-p_2) 
{\hat V}^{\rm eff}_{Z\ell_k\ell_m} \gamma_{\lambda} P_L v_{\ell_m}(p_3) \,,
\label{ampveff}
\end{equation}
 and  
\be
\Gamma(Z \to \ell_k \bar{\ell}_m) \simeq \frac{m_Z}{24\pi}|{\hat V}^{\rm eff}_{Z\ell_k\ell_m}|^{2},
\ee
with,
 \begin{align}
 {\hat V}^{\rm eff}_{Z\ell_k\ell_m}
&= \frac{1}{16 \pi^{2}} \frac{g}{c_W} \left(Y^{}_{\nu} Y_{\nu}^{\dagger} \right)^{km} 
\bigg\{ \left(\frac{1}{2}-s_{W}^{2}\right)B_1({\bf M_R},m_W) +(1-2s_{W}^{2}){\hat C}_{00}(p_2,p_1,{\bf M_R},m_W,m_W)   \nonumber\\
&  +2m_{W}^{2}{\hat C}_{0}(p_2,p_1,{\bf M_R},m_W,m_W) +\left(1-2s_{W}^{2}\right)m_{W}^{2}C_{2}(0,{\bf M_R},m_W)    \nonumber\\
&  +m_{W}^{2}\big( -2{\hat D}_{00}+m_{Z}^{2}({\hat D}_0+{\hat D}_1+{\hat D}_{13}-{\hat D}_{33})\big)(p_{2},0,p_{1},m_{W},0,{\bf M_R},0)   \nonumber\\
&  +m_{W}^{2}\big( -2{\hat D}_{00}+m_{Z}^{2}({\hat D}_0+{\hat D}_1+{\hat D}_{12}-{\hat D}_{22})\big)(p_{2},p_{1},0,m_{W},0,{\bf M_R},0)    \nonumber\\
&  +2c_{W}^{2}m_{W}^{2}(2{\hat D}_{00}-m_{Z}^{2}{\hat D}_{2})(0,p_{2},p_{1},0,{\bf M_R},m_{W},m_{W})  \bigg\} \nonumber\\
&+ \frac{1}{16 \pi^{2}} \frac{g}{c_W} v^{2} \left(Y^{}_{\nu} Y_{\nu}^{\dagger} Y^{}_{\nu} Y_{\nu}^{\dagger} \right)^{km} \bigg\{ \frac{1}{2}
  {\hat C}_{0}(p_{2},p_{1},m_W,{\bf M_R},{\bf M_R})   \nonumber\\
&  + \left( \frac{1}{2}-s_{W}^{2} \right) \left( 2{\hat D}_{00}(0,p_{2},p_{1},{\bf M_R},{\bf M_R},m_W,m_W) + C_2({\bf M_R},{\bf M_R},m_W) \right) \bigg\}\,. 
\label{Veffgeneric}
\end{align}
Here we have used again the notation in boldface for $M_R$, to mean that all these functions  have been expanded at large $M_R \gg v$, and we have kept just the first terms in these expansions. Specifically, we select all the needed terms in the involved loop functions that lead to contributions in $F_L^{\rm MIA}$ of  $\O(v^2/M_R^2)$, which are the first order terms in this large $M_R$ expansion of the form factor. For shortness,     
we leave the technical details of the loop functions expansions for the Appendix~\ref{Expansions}, and present here just the final result for the effective vertex. 
By plugging the results of Appendix~\ref{Expansions} into Eq.~\eqref{Veffgeneric}, we finally get:  
\begin{align}
\hspace{-.2cm}{\hat V}^{\rm eff}_{Z\ell_k\ell_m}  
&= \frac{g}{16 \pi^2 c_W} \left[
 \frac{m_W^2}{M_R^2} \left(f(c_W^2) + g(c_W^2) \log \left( \frac{m_W^2}{M_R^2} \right) \right)
 \left(Y^{}_{\nu} Y_{\nu}^{\dagger} \right)^{km}
 - \frac{v^2}{2M_R^2} 
\left( Y^{}_{\nu} Y_{\nu}^{\dagger}Y^{}_{\nu} Y_{\nu}^{\dagger} \right)^{km}  
\right] \nonumber \\
&= \frac{g}{16 \pi^2 c_W} \left[
 \frac{m_W^2}{M_R^2} \left(4.1+ i\, 2.1+1.4  \log \left( \frac{m_W^2}{M_R^2} \right) \right)
 \left(Y^{}_{\nu} Y_{\nu}^{\dagger} \right)^{km}
 - \frac{v^2}{2M_R^2} 
\left( Y^{}_{\nu} Y_{\nu}^{\dagger}Y^{}_{\nu} Y_{\nu}^{\dagger} \right)^{km}  
\right]\,,\label{veffsimple}
\end{align}
where, for practical purposes, the resulting coefficients $f(c_W^2)$ and $g(c_W^2)$, which are functions of the squared $W$ and $Z$ mass ratio, 
$m_W^2/m_Z^2= \cos^2 \theta_W \equiv c_W^2$, have been evaluated numerically in the second line, for $c_W^2=0.77$. Their complete analytical expressions are given by: 
\begin{align}
f(c_W^2)
&= 8(c_W^2+2)c_W^4 \arctan^2\Big[(4c_W^2-1)^{-\frac{1}{2}}\Big]
   +\frac{1}{6}\Big(-11 c_W^2 +12 c_W^4+4\pi ^2 (c_W^2+1)^2-26 \Big) \nonumber\\
& +\frac{1}{6}\Big(c_W^{-2}+18-28c_W^2-24c_W^4\Big) 
(4c_W^2-1)^{\frac{1}{2}}\arctan\Big[(4c_W^2-1)^{-\frac{1}{2}}\Big]+
\frac{1}{72}(10 -5c_W^{-2}) \nonumber\\
& -\log (c_W^2) \Big(3+2 c_W^2+(c_W^2+1)^2 \log (c_W^2)\Big) -2 (c_W^2+1)^2 \text{Li}_2(c_W^2+1)
    \nonumber\\
& -i \pi \Big( 3 +2 c_W^2+2 (c_W^2+1)^2 \log (c_W^2) \Big) \,,       \\
g(c_W^2)
&=
   \frac{1}{12} (c_W^{-2} +16)\,. 
\end{align}
Notice that in this on-shell $Z$ boson effective vertex there is an imaginary contribution. We have checked that it comes from diagrams (2a) and (2b) in Fig.~\ref{vertexdiags}, and is due to the possible crossing through the physical threshold of producing two light neutrinos (mainly $\nu_L$) from the on-shell $Z$ boson.
We have also checked that our analytical result for this imaginary part is in agreement with the analytical result of the limit of heavy singlets in 
reference~\cite{Ilakovac:1994kj}, i.e. for $m_{n_i}\gg m_W$.

\begin{figure}[t!]
\begin{center}
\includegraphics[width=.49\textwidth]{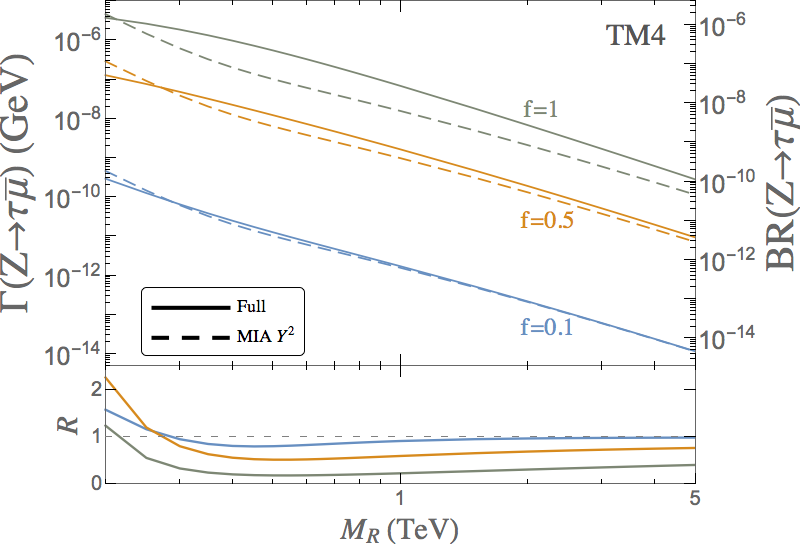} \hspace{0.05cm}
\includegraphics[width=.49\textwidth]{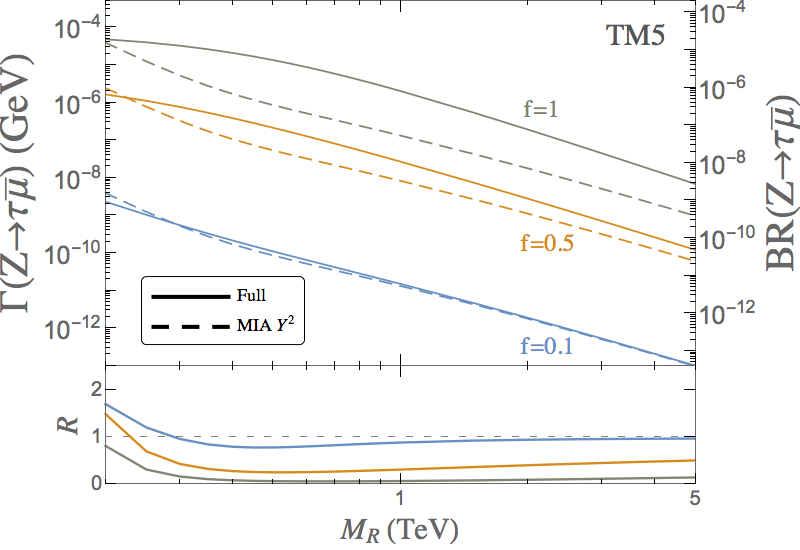}
\caption{Predictions for the partial width $\Gamma (Z \to \tau \bar \mu)$ and branching ratio BR$(Z \to \tau \bar \mu)$ as a function of $M_R$.  The dashed lines are the predictions from the MIA to ${\cal O}(Y_\nu^2)$. The solid lines are the predictions from the full one-loop computation of the mass basis. Here the examples TM4 (left panel) and TM5 (right panel) with f=0.1,0.5,1, as explained in the text,  are chosen. In the bottom of these plots the ratio $R=\Gamma_{\rm MIA}/ \Gamma_{\rm full}$ is also shown. }\label{totalcomparisonY2}
\end{center}
\end{figure}

\begin{figure}[t!]
\begin{center}
\includegraphics[width=.49\textwidth]{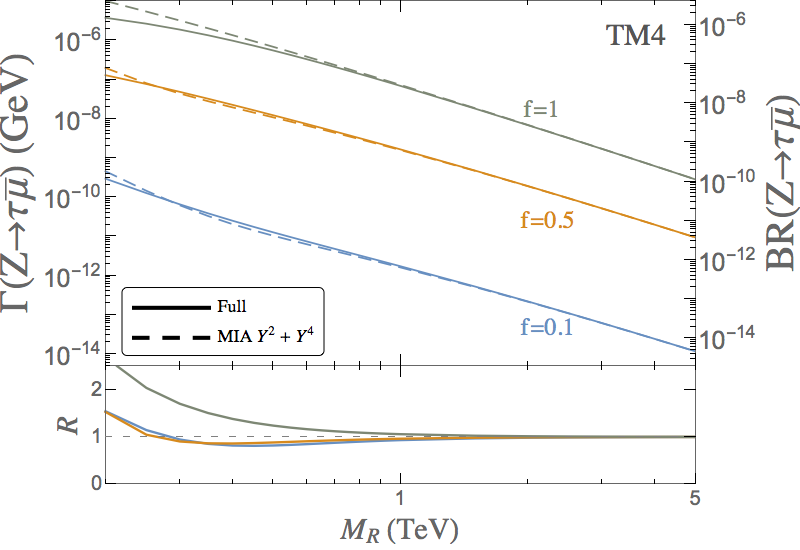}\hspace{0.05cm}
\includegraphics[width=.49\textwidth]{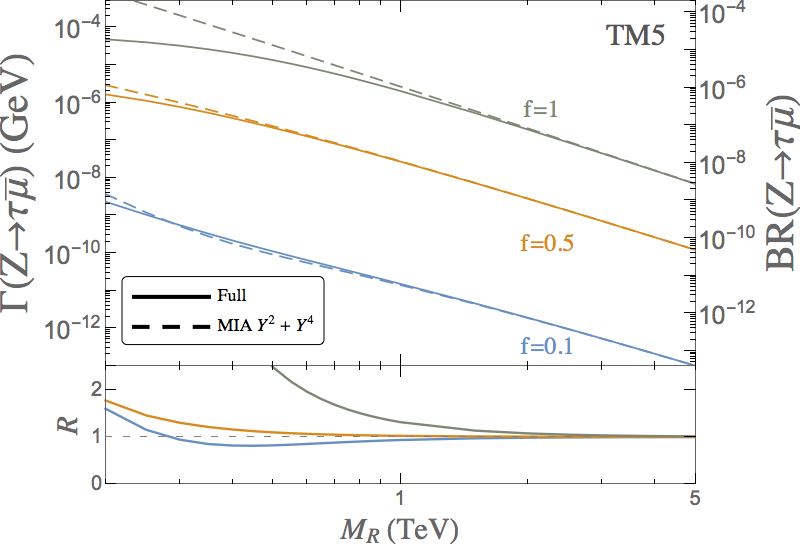}
\caption{Predictions for the partial width $\Gamma (Z \to \tau \bar \mu)$ and branching ratio BR$(Z \to \tau \bar \mu)$ as a function of $M_R$.  The dashed lines are the predictions from the MIA to ${\cal O}(Y_\nu^2+Y_\nu^4)$. The solid lines are the predictions from the full-one loop computation of the mass basis. Here the examples TM4 (left panel) and TM5 (right panel) with f=0.1,0.5,1, as explained in the text,  are chosen. In the bottom of these plots the ratio $R=\Gamma_{\rm MIA}/ \Gamma_{\rm full}$ is also shown. }\label{totalcomparisonY2Y4}
\end{center}
\end{figure}

In Figs.~\ref{totalcomparisonY2} and \ref{totalcomparisonY2Y4} we present the numerical results for our predictions of the partial widths and the corresponding branching ratios for the LFVZD. For illustrative purposes we have chosen two examples of input neutrino Yukawa coupling matrices, following \cite{Arganda:2014dta, Arganda:2015ija, Arganda:2015naa,DeRomeri:2016gum}. Concretely, in these figures we use:
\begin{align}
Y_{\nu}^{\rm TM4}=f\left(\begin{array}{ccc}
0.1&0&0\\0&1&0\\0&1&0.014
\end{array}\right)\,, &\quad
Y_{\nu}^{\rm TM5}=f\left(\begin{array}{ccc}
0&1&-1\\0.9&1&1\\1&1&1
\end{array}\right) \,.
\label{YnuTM}
\end{align}
  These particular textures  were selected as illustrative examples belonging to a type of scenarios (named TM scenarios in \cite{DeRomeri:2016gum}) in which the LFV is always extremely suppressed in the $\mu e$ sector, which is well-known to be  highly constrained,  but it can lead to large LFV in  the $\tau \mu$ sector, which is less severely constrained. These scenarios are known to produce interesting phenomenological implications. For  instance,  in collider physics they can lead to  the production of exotic   $\tau$-$\mu$-jet-jet events at LHC~\cite{Arganda:2015ija}. Notice that
the Yukawa coupling matrices in these examples are usually given in terms of a  scaling factor $f$ that characterizes the global strength of the coupling. We have also tried other examples of textures leading instead to large LFV in  the $\tau e$ sector (the so-called TE scenarios in~\cite{DeRomeri:2016gum}) and the results are quite similar to the ones presented here for the $\tau \mu$ sector. Since our aim in this section is mainly to provide useful and accurate formulas for the LFV effective vertex, we believe that the choice of these two textures should be sufficient for the check of the $\hat V^{\rm eff}_{Z\ell_k\ell_m}$ accuracy.   
It should also be noted that in checking our numerical results from the MIA with the previous full one-loop numerical results of~\cite{DeRomeri:2016gum}, which is our choice here,  we are also simultaneously checking the agreement with all the other numerical results of  the physical basis in~\cite{Ilakovac:1994kj,Illana:2000ic,Abada:2014cca} since they are all consistent.

In Figs.~\ref{totalcomparisonY2} and \ref{totalcomparisonY2Y4} we have compared the MIA results with the full results in order to learn on the goodness of our approximate formulas for the on-shell effective vertex. We show the results both to 
${\cal O}(Y_\nu^2)$, Fig.~\ref{totalcomparisonY2}, and to ${\cal O}(Y_\nu^2+Y_\nu^4)$, Fig.~\ref{totalcomparisonY2Y4}, for comparison. The first thing worth noticing is that with  the LO result, i.e. taking just the ${\cal O}(Y_\nu^2)$ solution for the effective vertex, the agreement between the MIA and the full result is not so good as taking also the NLO terms of ${\cal O}(Y_\nu^4)$. It is also clear from these plots that our simple formula in Eq.~\eqref{veffsimple} for the MIA effective vertex to ${\cal O}(Y_\nu^2+Y_\nu^4)$ provides very accurate results, leading to LFV rates which are very close to the full results, even for large Yukawa couplings, with global strength $f$ of order 1. It is only for relatively low values of $M_R$, say well below 1 TeV, and very large Yukawa couplings, say with $f \geq {\cal O}(1)$, where the initial assumption of $m_D\ll M_R$ does not hold anymore, that we get a significant deviation from the full results. For all the other input parameters the agreement is excellent. Therefore the MIA approximation works pretty well in the present case of LFVZD.

Finally, to end this section we find interesting to compare our result for 
${\hat V}^{\rm eff}_{Z\ell_k\ell_m}$ in Eq.~\eqref{veffsimple} with our previous result in \cite{Arganda:2017vdb} for the corresponding on-shell Higgs effective vertex  ${\hat V}^{\rm eff}_{H\ell_k\ell_m}$ which is the proper one for the LFV Higgs decays 
$H \to \ell_k \bar{\ell}_m$. This Higgs effective vertex computed in \cite{Arganda:2017vdb} was obtained in exactly the same context of ISS with heavy right handed neutrinos and following the same MIA and large $M_R$ techniques as in the present paper, therefore this comparison gives us a valuable information. In the Higgs case, the amplitude and partial decay width can be written as:
\begin{equation}
i {\cal M} \simeq -i \bar{u}_{\ell_k} {\hat V}_{H \ell_{k}\ell_{m}}^{\rm eff} P_L  v_{\ell_m} \, , 
\label{amplitudeVeff}
\end{equation}
\be
\Gamma (H \to \ell_{k}\bar{\ell}_{m})\simeq \frac{m_H}{16\pi} \big\vert 
{\hat V}_{H \ell_{k}\ell_{m}}^{\rm eff}\big\vert^{2} \, ,
\label{Gammaeff}
\ee
and the on-shell vertex, i.e. for $p_1^2=m_H^2$, found is \cite{Arganda:2017vdb}: 
\be
{\hat V}_{H \ell_{k} \ell_{m}}^{\rm eff}=\frac{g}{64 \pi^{2}} \frac{m_{\ell_k}}{m_{W}}  \left[  \frac{m_{H}^{2}}{M_{R}^{2}}
\left( r\Big(\frac{m_{W}^{2}}{m_{H}^{2}}\Big) +\log\left(\frac{m_{W}^{2}}{M_{R}^{2}}\right) \right) \left(Y^{}_{\nu} Y_{\nu}^{\dagger}\right)^{km} - \frac{3v^{2}} {M_{R}^{2}} \left(Y^{}_{\nu} Y_{\nu}^{\dagger} Y^{}_{\nu} Y_{\nu}^{\dagger} \right)^{km} \right]\,,
\label{Veffsimple} 
\ee 
with a numerical value for the coefficient given by $r(m_W^2/m_H^2)\simeq 0.3$.
We clearly see that the two effective vertices ${\hat V}_{Z \ell_{k} \ell_{m}}^{\rm eff}$ and ${\hat V}_{H \ell_{k} \ell_{m}}^{\rm eff}$ have the same functional form as functions of the input parameters $M_R$ and $Y_\nu$. Both show a decoupling behavior with the heavy neutrino masses as $\sim 1/M_R^2$, both have a term with $\log(m_W^2/M_R^2)$, and both have contributions from the LO, ${\cal O}(Y_\nu^2)$,  and from the NLO, ${\cal O} (Y_\nu^4)$.   Besides, we have checked that both contributions LO and NLO in the two decays, $H$ and $Z$, are needed to get a good numerical agreement of the MIA with the full one-loop result.
The main difference, therefore, is just the numerical values of the coefficients in front of these terms. 
First, in the Higgs case the on-shell effective vertex is real. There is not an imaginary part because, contrary to the $Z$ decays, there are not diagrams with two light neutrino lines (mainly $\nu_L$) from the Higgs boson.
Instead, the two neutrino lines connected to a Higgs particle are one light (mainly $\nu_L$) and the other one heavy (mainly $\nu_R$) that cannot be produced on-shell in the Higgs decay, under our assumption of heavy $M_R \gg v$.  
Second, notice that the $Z$ effective vertex, ${\hat V}^{\rm eff}_{Z\ell_k \ell_m}$, is universal in flavor, namely, its size does not depend on the mass of the charged leptons involved, $m_{\ell_{k,m}}$. This is also in contrast with the Higgs case, where ${\hat V}^{\rm eff}_{H\ell_k\ell_m}$ shows a linear dependence with the heaviest charged lepton mass, $m_{\ell_{k}}$, indicating a larger LFV effect for heavier charged leptons.      

\section{Numerical estimates with the MIA-effective vertex of maximum allowed LFVZD rates}
\label{maximumLFVZD}

In order to show the applicability and simplicity of the MIA results, in this section we use the effective vertex in Eq.~\eqref{veffsimple} to compute the maximum LFVZD rates in the ISS model that are allowed by present experimental constraints. 
For that purpose, we use the constraints derived in~\cite{Fernandez-Martinez:2016lgt}, 
where the non-unitary matrix $N$ describing the mixing between the light neutrino mass eigenstates and the SM charged leptons via $W$ interactions was parametrized in terms of a small Hermitian matrix $\eta$ defined by~\cite{Fernandez-Martinez:2016lgt}: 
\begin{equation}
N = (1-\eta) U_{\rm PMNS}\,.
\end{equation}
This $\eta$ matrix then encodes the deviations from the unitary $U_{\rm PMNS}$ induced by the mixing with the extra heavy neutrinos.

By performing a global fit analysis, upper bounds\footnote{Notice that we have corrected a typo in the $\eta_{3\sigma}^{\rm max}$ given in~\cite{Arganda:2017vdb}, which was present only in the text, not in the codes. } 
on the $\eta$ matrix were set to
\begin{align}
 \eta_{3\sigma}^{\rm max}=\left(\begin{array}{ccc}
1.62\times 10^{-3}&1.51\times 10^{-5}&1.57\times 10^{-3}\\1.51\times 10^{-5}&3.92\times 10^{-4}&9.24\times 10^{-4}
\\1.57\times 10^{-3}&9.24\times 10^{-4}&3.67\times 10^{-3}
\end{array}\right)\,.
\label{etamax3sigma}
\end{align}
In our case of interest with degenerate $M_R$ and $v Y_\nu\ll M_R$, the $\eta$ matrix can be written, following ~\cite{Fernandez-Martinez:2016lgt}, approximately as: 
\begin{equation}\label{etaY2}
\eta=\frac{v^2}{2M_R^2}\, Y_\nu^{} Y_\nu^\dagger\,,
\end{equation}
what allows us to define a scenario that easily implements these bounds.
Following \cite{Arganda:2017vdb}, we define this scenario by considering the following neutrino Yukawa matrix:
 \begin{align}
 Y_{\nu}^{\rm GF}=f\left(\begin{array}{ccc}
 0.33&0.83&0.6\\-0.5&0.13&0.1\\-0.87&1&1
 \end{array}\right)\,.
 \label{YnuGF}
\end{align}
This $Y_\nu$ leads to a $Y_\nu Y_\nu^\dagger$ with the same pattern as in Eq.~\eqref{etamax3sigma} and
 saturates the $\eta_{3\sigma}^{\rm max}$ bounds for $f/M_R= (3/10)\, {\rm TeV}^{-1}$.
Consequently, it provides a simple way for concluding on maximum allowed rates within this model. 
Notice that one can always take $Y_\nu$ and $M_R$ as independent input parameters as long as $\mu_X$ accommodates light neutrino oscillation data by means of the $\mu_X$-parametrization introduced in~\cite{Arganda:2014dta}. 

\begin{figure}[t!]
\begin{center}
\includegraphics[width=.49\textwidth]{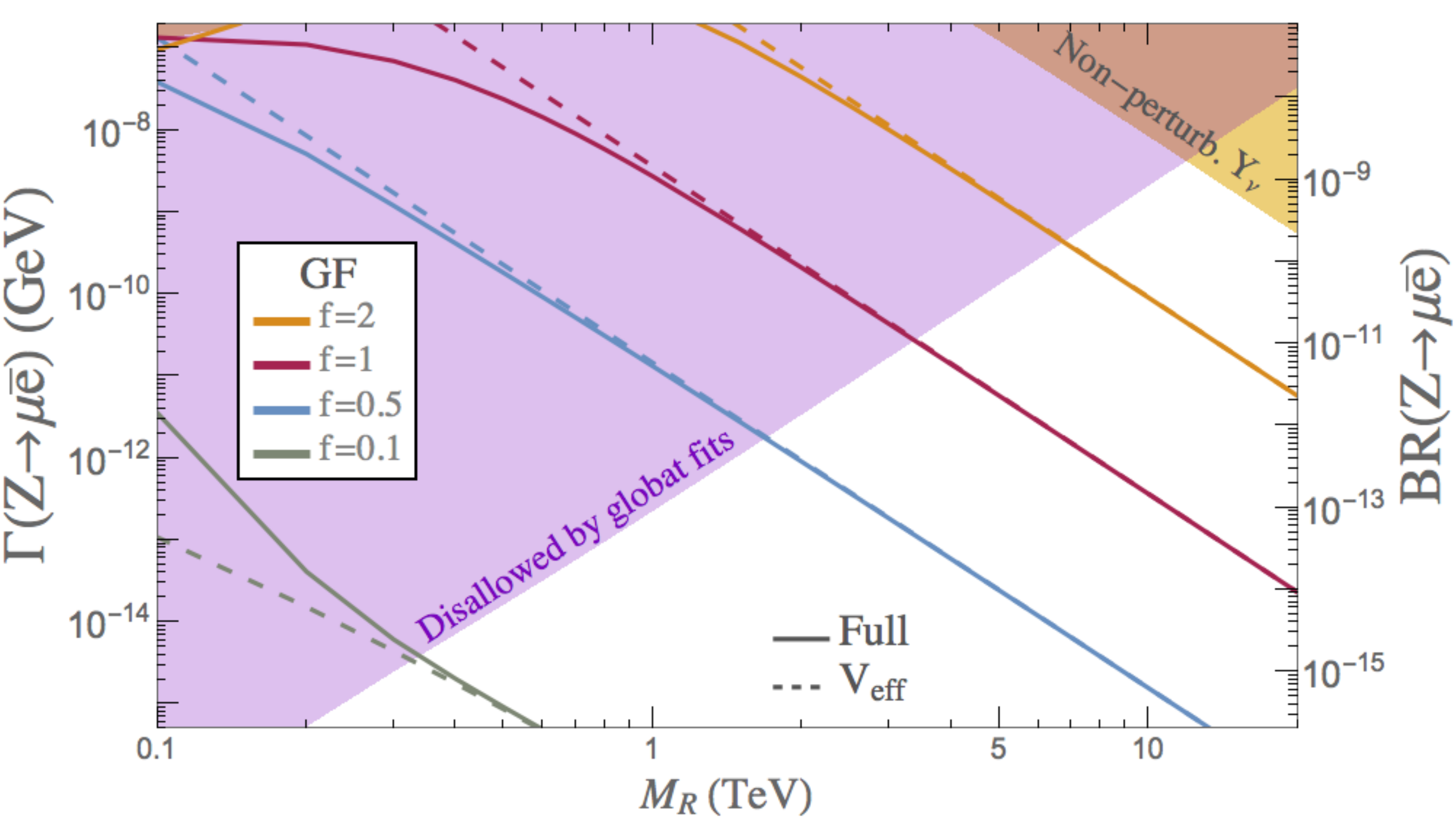}\\
\includegraphics[width=.49\textwidth]{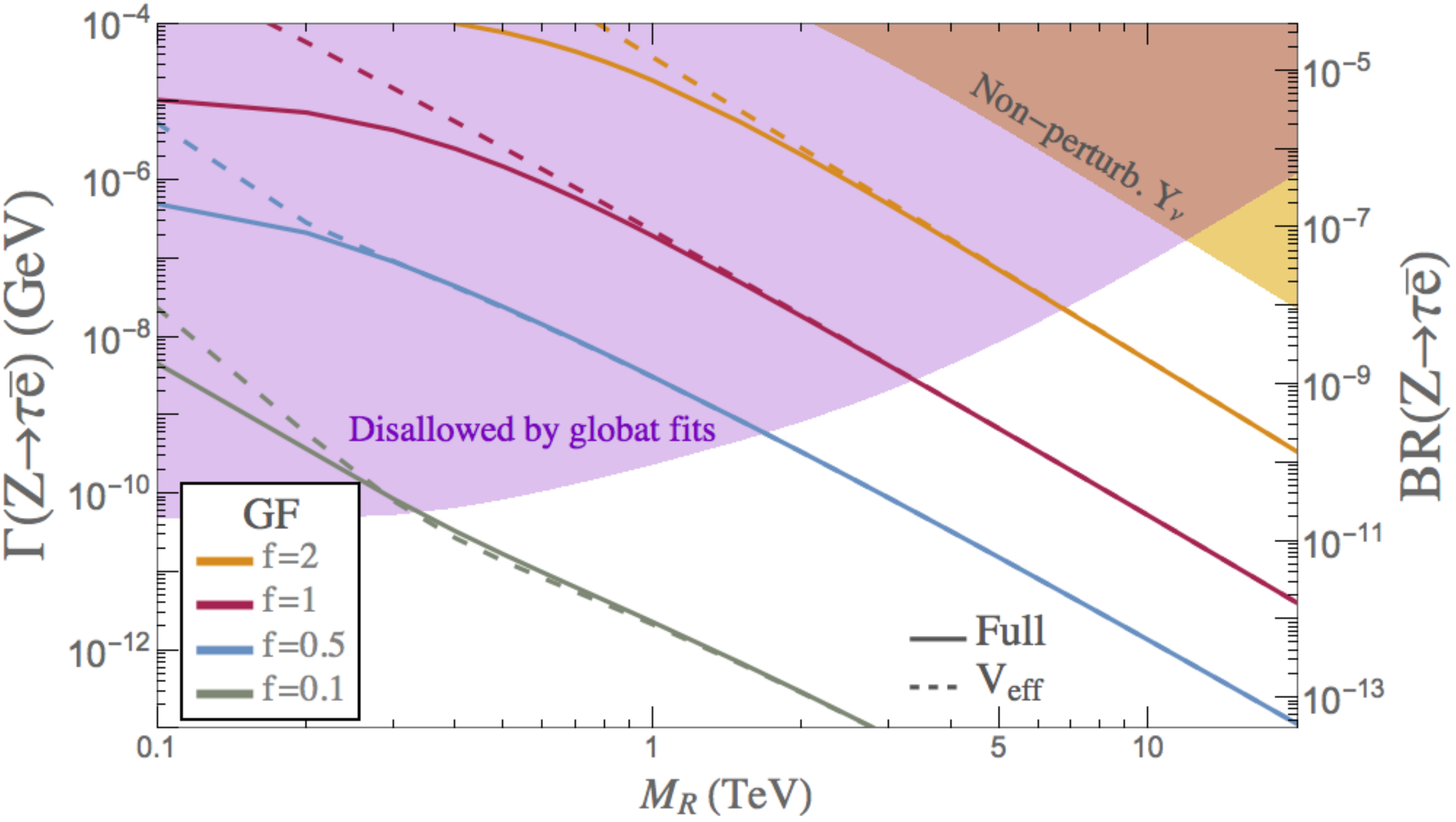}
\hspace{0.05cm}
\includegraphics[width=.49\textwidth]{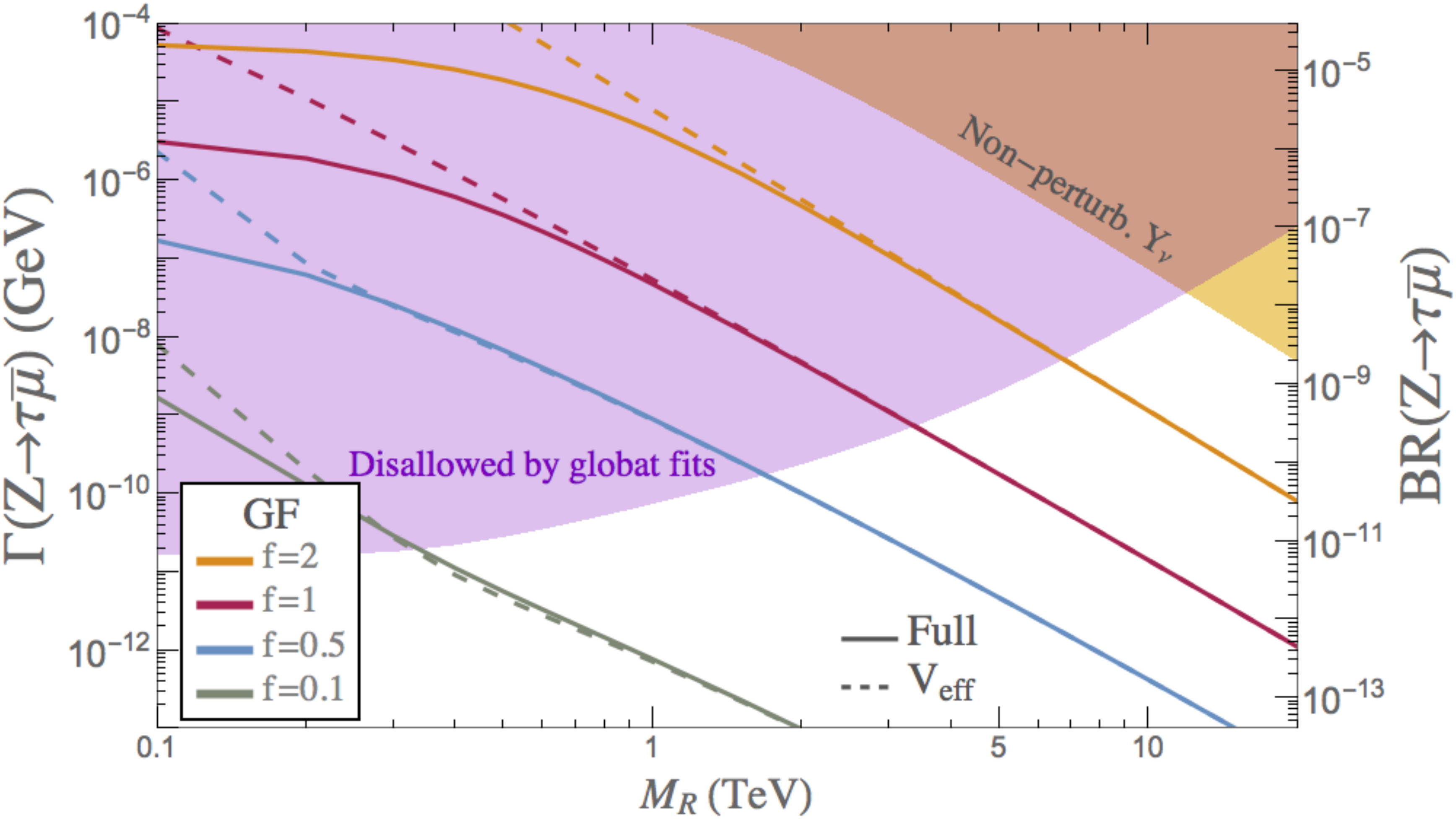}
\caption{Predictions for $Z\to\mu \bar e$ (top), $Z \to \tau  \bar e$ (left) and 
$Z \to \tau  \bar\mu$ (right) using the effective vertex computed with the MIA (dashed lines) and the full results in the mass basis (solid lines) for $Y_\nu^{\rm GF}$ in Eq.~\eqref{YnuGF} and $f=0.1, 0.5, 1, 2$. The chosen example GF is explained in the text.
Shadowed areas are disallowed by global fit results (purple) or for giving non-perturbative Yukawa couplings (yellow).}
\label{LFVZD_GF}
\end{center}
\end{figure}

We show in Fig.~\ref{LFVZD_GF} the results for the three LFVZD channels in the GF scenario defined in Eq.~\eqref{YnuGF}.
Solid lines are the exact one-loop results, computed with the expressions in Appendix~\ref{app:Full} after diagonalizing to the mass basis, while dashed lines have been obtained using the effective vertex in Eq.~(\ref{veffsimple}) in terms of the parameters in the EW basis.
Shadowed areas represent the regions disallowed by some of the constraints:
in the purple area, covering the upper and left parts of the figures, the upper bounds in Eq.~\eqref{etamax3sigma} are not fulfilled; 
in the yellow area, in the upper right corners, the Yukawa coupling matrix  becomes non-perturbative.
Our criteria for perturbativity is imposing $|Y_{ij}|^2/4\pi<1$ for all the entries, what implies $f<\sqrt{4\pi}$ for this scenario. 

The first thing we conclude from these plots is that the computed effective vertex works extremely well in the allowed white region. 
Applying the constraints from both global fits and perturbativity imposes an upper bound on $v Y_\nu/M_R$, which further supports our criteria of not computing higher order terms in Eq.~\eqref{FLtot_FtH}.
For masses below the TeV scale, when $M_R$ is close to the EW scale, the assumption $v/M_R\ll 1$ breaks down and the effective vertex stops being a good approximation. 
Nevertheless, from these plots we see that the MIA results work very well in the allowed region also for lighter $M_R$.
Consequently, we can conclude that our effective vertex is a very powerful tool to easily estimate the LFVZD rates in the region of $M_R\gtrsim300$ GeV that is allowed by present constraints.

Second, we  see that the shape of the excluded purple area, or the complementary allowed white area, is different in the $\mu e$ sector with respect to the $\tau e$ and $\tau \mu$ ones, specially in the low $f$ and low $M_R$ regime.
The origin of this difference comes from the strong bound on $\eta_{e\mu}$, coming from the upper bound on $\mu\to e\gamma$ by MEG~\cite{TheMEG:2016wtm}, which suppresses the $\O(Y_\nu^2)$ contributions that are the most relevant ones at this low $f$ regime.

Finally, we can use Fig.~\ref{LFVZD_GF} to conclude on the maximum allowed rates for the LFVZD. 
As it happens for the Higgs case~\cite{Arganda:2017vdb}, these large rates are found in the crossing between the global fit and perturbativity bounds, which happens  at  heavy masses around 10 TeV. 
Taking the benchmark sensitivities of $10^{-9}$ for the future linear colliders and assuming a modest improvement in the sensitivities of $10^{-11}$ at FCC-ee, we see that these rates could be accessible at both experiments for the three LFVZD channels. This is in contrast to the $H$ decays, where the $H\to \mu e$ channel is further suppressed due to the small lepton masses. The difference between the $Z$ and $H$ decays comes from the the {\it flavor universality} in the LFVZD, as we discussed before.

Interestingly, these future experiments could access not only the high $M_R$ regime, but also  lighter values, meaning that they could be complementary to direct searches at the LHC (for a recent summary, see for instance~\cite{Cai:2017mow}).
Indeed, the FCC-ee could be able to explore the full allowed mass range from the EW scale up to masses above the TeV scale in the $\tau e$ and $\tau \mu$ sectors.

Summarizing our findings, we conclude that our effective vertex provides a simple and useful tool for estimating the LFVZD rates in the allowed region, which can be as large as,
\begin{align}
{\rm BR}(Z\to \mu e) \lesssim&~ 10^{-9}~\big(10^{-13}\big) \qquad {\rm for}~M_R\sim10~{\rm TeV}~(500~{\rm  GeV})\,, \\
{\rm BR}(Z\to \tau e ) \lesssim&~ 10^{-7}~\big(10^{-10}\big) \qquad {\rm for}~M_R\sim10~{\rm TeV}~(500~{\rm  GeV})\,, \\
{\rm BR}(Z\to \tau\mu) \lesssim&~ 10^{-8}~\big(10^{-11}\big) \qquad {\rm for}~M_R\sim10~{\rm TeV}~(500~{\rm  GeV})\,,
\end{align}
implying that future lepton colliders could probe this kind of low scale seesaw models looking for LFVZD. 

\section{Conclusions}
\label{conclusions}
In this work we have studied the lepton flavor violating decays of the $Z$ boson into two leptons with different flavor. We have computed in full detail the one-loop contributions from the heavy right handed neutrinos to these decays within the inverse seesaw and by using the mass insertion approximation, which works with the electroweak neutrino basis, instead of the usual full one-loop computation that works with the    neutrino mass basis. Our analytical results of the involved form factors from the mass insertion approximation  are presented explicitly in terms of the relevant inverse seesaw parameters: the right handed neutrino mass, $M_R$, and the neutrino Yukawa coupling matrix, $Y_\nu$. The formulas presented here are simple and useful. They contain the LO contributions of ${\cal O}(Y_\nu^2)$ and the NLO contributions of  ${\cal O}(Y_\nu^4)$, both being relevant for the kind of scenarios that we are interested in with large neutrino Yukawa couplings, $Y_\nu \sim {\cal O}(1)$.
We have then presented our computation of the one-loop effective vertex $Z\ell_k\ell_m$ which is derived from the large $M_R$ expansion, valid for $M_R \gg v$,  of the form factors and by keeping the first order in this expansion which turns out to be of ${\cal O}(v^2/M_R^2)$. This demonstrates explicitly the decoupling behavior of the heavy right handed neutrinos. 

As a very important test of our analytical results,  the work has been completed with an explicit demonstration of the gauge invariance of our results for the on-shell effective one-loop vertex. 
   
In the last part we have applied this effective vertex for an easy and accurate estimate of  the  maximum allowed lepton flavor violating $Z$ decay rates by present data in these low scale seesaw models. The rates found are indeed  promising, since they are at the reach of future lepton colliders.

\section*{Acknowledgments}
We warmly thank our colleague and friend Ernesto Arganda for his valuable collaboration in our long standing project on LFV phenomenology. We also acknowledge his participation in the preliminary discussions that triggered this   present research work.    
This work is supported by the European Union through the ITN ELUSIVES H2020-MSCA-ITN-2015//674896 and the RISE INVISIBLESPLUS H2020-MSCA-RISE-2015//690575, by the CICYT through the project FPA2016-78645-P,  and by the Spanish MINECO's ``Centro de Excelencia Severo Ochoa''  Programme under grant SEV-2016-0597. This work was partially supported by ANPCyT PICT 2013-2266 and PICT 2016-0164 (A.S. and R.M.).


\begin{appendix}
\section{Appendix: Conventions for the one-loop integrals}
\label{LoopIntegrals}
 In all this work, we use the following definitions and conventions for the one-loop integrals and the involved momenta:  
\begin{equation}\label{loopfunctionB}
    \mu^{4-d}~ \int \frac{d^d k}{(2\pi)^d} \frac{\{1; k^{\mu}\}}
    {[k^2 - m_1^2][(k+p_1)^2 - m_2^2]}
    = \frac{i}{16 \pi^2} \left\{ B_0; B^{\mu} \right\}
    (p_1,m_1,m_2)\,,
\end{equation}
\begin{align}
    \mu^{4-d}& \int \frac{d^d k}{(2\pi)^d}
    \frac{\{1;   k^{\mu}; k^{\mu}k^{\nu}\}}
    {[k^2 - m_1^2][(k + p_1)^2 - m_2^2][(k + p_1 + p_2)^2 - m_3^2]}
    \nonumber \\
     & =  \frac{i}{16 \pi^2}
    \left\{ C_0;   C^{\mu}; C^{\mu \nu} \right\}
    (p_1, p_2, m_1, m_2, m_3)\,, \label{loopfunctionC}
\end{align}
\begin{align}
   \mu^{4-d}& \int \frac{d^d k}{(2\pi)^d}
    \frac{\{1;   k^{\mu}; k^{\mu}k^{\nu}\}}
    {[k^2 - m_1^2][(k + p_1)^2 - m_2^2][(k + p_1 + p_2)^2 - m_3^2][(k + p_1 + p_2+ p_3)^2 - m_4^2]}
    \nonumber \\
     & =  \frac{i}{16 \pi^2}
    \left\{ D_0;   D^{\mu}; D^{\mu \nu}\right\}
    (p_1, p_2, p_3, m_1, m_2, m_3,m_4)\,. \label{loopfunctionD}
\end{align}

In terms of momenta, the decompositions are:
\begin{align}
B^{\mu}(p_1,m_1,m_2) &= p_{1}^{\mu}B_1(p_1,m_1,m_2)  \nonumber\\
C^{\mu}(p_1, p_2, m_1, m_2, m_3) &= \big\{ p_{1}^{\mu}C_1 +p_{2}^{\mu}C_2 \big\}(p_1, p_2, m_1, m_2, m_3)  \nonumber\\
C^{\mu\nu}(p_1, p_2, m_1, m_2, m_3) &= \big\{ g^{\mu\nu}C_{00} +p_{1}^{\mu}p_{1}^{\nu}C_{11} +p_{1}^{\mu}p_{2}^{\nu}C_{12} \nonumber\\
&\hspace{3mm}+p_{2}^{\mu}p_{1}^{\nu}C_{21} +p_{2}^{\mu}p_{2}^{\nu}C_{22} \big\}(p_1, p_2, m_1, m_2, m_3)  \nonumber\\
D^{\mu}(p_1, p_2,p_3, m_1, m_2, m_3, m_4) &= \big\{ p_{1}^{\mu}D_1 +p_{2}^{\mu}D_2 +p_{3}^{\mu}D_3 \big\}(p_1, p_2, p_3, m_1, m_2, m_3, m_4)  \nonumber\\
D^{\mu\nu}(p_1, p_2,p_3, m_1, m_2, m_3, m_4) &= \big\{ g^{\mu\nu}D_{00} +p_{1}^{\mu}p_{1}^{\nu}D_{11} +p_{1}^{\mu}p_{2}^{\nu}D_{12} +p_{1}^{\mu}p_{3}^{\nu}D_{13}     \nonumber\\
& \hspace{3mm} +p_{2}^{\mu}p_{1}^{\nu}D_{21} +p_{2}^{\mu}p_{2}^{\nu}D_{22}+p_{2}^{\mu}p_{3}^{\nu}D_{23} +p_{3}^{\mu}p_{1}^{\nu}D_{31} \nonumber\\
& \hspace{3mm}+p_{3}^{\mu}p_{2}^{\nu}D_{32} +p_{3}^{\mu}p_{3}^{\nu}D_{33} \big\}(p_1, p_2, p_3, m_1, m_2, m_3, m_4) \,.
\end{align}
We adopt the usual definitions in dimensional regularization, with:
\be\label{DeltaDiv}
\Delta= 2/\epsilon-\gamma_E +\log(4\pi),
\ee 
and $d=4-\epsilon$. We name $\mu$ the usual regularization scale. 

In the following we use a shorten notation for the loop functions when evaluated at zero external momenta:
\begin{align}
B(m_1,m_2) &\equiv B(p_1,m_1,m_2)|_{p_1^2=0}  \,,\nonumber\\
C(m_1, m_2, m_3) &\equiv C(p_1, p_2, m_1, m_2, m_3)|_{p_1^2=p_2^2=0}  \,,\nonumber\\
D(m_1, m_2, m_3, m_4) &\equiv D(p_1, p_2, p_3, m_1, m_2, m_3, m_4)|_{p_1^2=p_2^2=p_3^2=0}\,.   
\end{align}
 We also use a shorten notation for the case when the loop functions are evaluated at on-shell $Z$ external momentum, i.e. for $p_1^2=m_Z^2$. Specifically, we refer to all these functions by using a hat notation. For instance: 
\be
{\hat C}_{0}(p_2,p_1,M_R,m_W,m_W)\equiv C_{0}(p_2,p_1,M_R,m_W,m_W)|_{p_1^2=m_Z^2},
\ee  
and similarly for the other functions with a hat appearing in the text. 
Since, we are neglecting the external fermion masses, we are also taking $p_2^2 = 0$ and $p_3^2 = 0$, though for shortness these are not explicitly written through the text.
\section{Appendix: Full Form Factors (Feynman-'t Hooft gauge)}
\label{app:Full}

\begin{figure}[t!]
\begin{center}
\includegraphics[width=\textwidth]{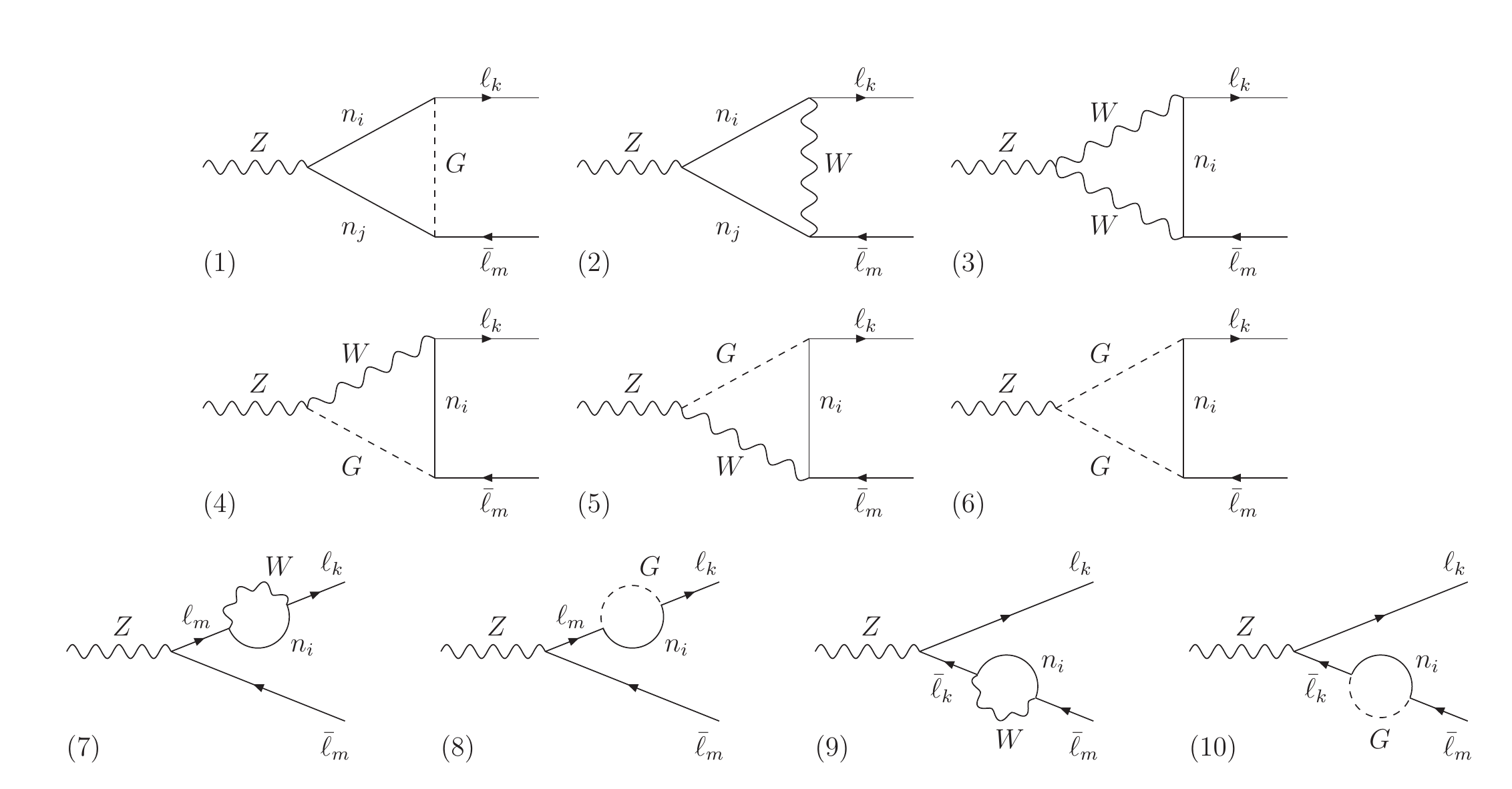}
\caption{One-loop diagrams in the Feynman-'t Hooft gauge  contributing to the full computation of $Z\to\ell_k {\bar \ell_m}$ decays in the physical neutrino mass eigenstate basis.}
\label{DiagramsMassbasis}
\end{center}
\end{figure}

For completeness, and to better clarify the comparison with our MIA computation, we include here the full form factors in the Feynman-'t Hooft gauge of the different diagrams in the neutrino physical basis.
We took the formulas from~\cite{Illana:2000ic} and rewrite them using the notation introduced in Appendix~\ref{LoopIntegrals}. 

\begin{equation}
\mathcal F_Z^{(1)} = \frac12\, B_{\ell_k n_i}^{} B^*_{\ell_m n_j}\left\{ - C_{n_i n_j} \,x_i x_j \, m_W^2 \hat C_0 + C_{n_i n_j}^* \sqrt{x_ix_j} \Big[m_Z^2\, \big(\hat  C_{12}-\hat C_{22}\big) - 2 \hat C_{00}+\frac12\Big]\right\},
\end{equation}
where $\hat C_{0, 00, 12, 22}\equiv \hat C_{0, 00, 12, 22}(p_2,p_1,m_W,m_{n_i}, m_{n_j})$;
\begin{equation}
\hspace{-.2cm}\mathcal F_Z^{(2)}  = B_{\ell_k n_i}^{}B^*_{\ell_m n_j}\left\{ -C_{n_i n_j} \Big[m_Z^2 \Big( \hat C_0 + \hat C_1 + \hat C_{12} -\hat C_{22} \Big) - 2 \hat C_{00} +1  \Big]+ C_{n_i n_j}^* \sqrt{x_ix_j} \, m_W^2 \hat C_0\right\},
\end{equation}
where $\hat C_{0, 1, 00, 12, 22}\equiv \hat C_{0, 1, 00, 12, 22}(p_2,p_1,m_W,m_{n_i}, m_{n_j})$;
\begin{equation}
\mathcal F_Z^{(3)} = 2c_W^2 B_{\ell_k n_i}^{} B^*_{\ell_m n_i}\left\{  m_Z^2 \Big(\hat C_1+\hat C_{12}-\hat C_{22}\Big)-6 \hat C_{00}+1\right\} ,
\end{equation}
where $\hat C_{1, 00,12, 22}\equiv \hat C_{1, 00,12, 22}(p_2,p_1,m_{n_i},m_W,m_W)$;
\begin{equation}
\mathcal F_Z^{(4)}+\mathcal F_Z^{(5)}  = -2s_W^2\, B_{\ell_k n_i}^{} B^*_{\ell_m n_i}  \, x_i\, m_W^2  \hat C_0 ,
\end{equation}
where $\hat C_{0}\equiv \hat C_{0}(p_2,p_1,m_{n_i},m_W,m_W)$;
\begin{equation}
\mathcal F_Z^{(6)} = -(1-2s_W^2)\,  B_{\ell_k n_i}^{} B^*_{\ell_m n_i}\,  x_i\, \hat C_{00},
\end{equation}
where  $\hat C_{00}\equiv \hat C_{00}(p_2,p_1,m_{n_i},m_W,m_W)$;
\begin{equation}
\mathcal F_Z^{(7)}+\mathcal F_Z^{(8)}+\mathcal F_Z^{(9)}+\mathcal F_Z^{(10)}  = \frac12(1-2c_W^2)\, B_{\ell_k n_i}^{} B^*_{\ell_m n_i}\left\{(2+x_i) B_1+1\right\},
\end{equation}
where $ B_1\equiv  B_1(m_{n_i},m_W)$.

In all these formulas,  sum over neutrino indices, $i,j=1,... ,9$ has to be understood and $x_i\equiv~m_{n_i}^2/m_W^2$. 
As before, the loop functions with a hat means that they are evaluated at on-shell external momenta, i.e. at $p_1^2=m_Z^2$, and $p_2^ 2=p_3^2=0$ since we are neglecting the charged lepton masses.

In the neutrino mass basis, the relevant couplings are given by the following terms in the Lagrangian:
\begin{align}
\mathcal L_W&=-\dfrac g{\sqrt{2}} \sum_{i=1}^{3}\sum_{j=1}^{9} W^-_\mu \bar\ell_i B_{\ell_i n_j} \gamma^\mu P_L n_j + h.c., \\
\mathcal L_Z &= -\dfrac g{4c_W} \sum_{i,j=1}^{9}Z_\mu\, \bar n_i \gamma^\mu \Big[C_{n_i n_j} P_L - C_{n_in_j}^* P_R\Big]n_j ,\\
\mathcal L_H &=-\dfrac g{2m_W}  \sum_{i,j=1}^{9} H\,\bar n_i C_{n_in_j}\Big[m_{n_i}P_L+m_{n_j} P_R\Big]n_j,\\
\mathcal{L}_{G^{\pm}} &= -\frac{g}{\sqrt{2} m_W}\sum_{i=1}^{3}\sum_{j=1}^{9} G^{-}\bar{\ell_i} B_{\ell_i n_j} \Big[m_{\ell_i} P_L - m_{n_j} P_R \Big]n_j  + h.c\,,\\ 
\mathcal{L}_{G^{0}} & =-\dfrac {ig}{2 m_W} \sum_{i,j=1}^{9}G^0\, \bar n_i  C_{n_in_j} \Big[m_{n_i}  P_L - m_{n_j} P_R\Big]n_j ,  
\end{align}
where $U^\nu$ is a unitary rotation matrix that diagonalizes the neutrino mass matrix, $M_{\rm ISS}$, according to, 
\be
U^{\nu^T}M_{\rm ISS}\, U^\nu = {\rm diag}(m_{n_1},\dots,m_{n_9})\,,
\ee
and
\begin{align}
\label{eq:BCmatrices}
B_{\ell_in_j}&=U_{ij}^{\nu*}, \\
C_{n_in_j}&=\sum_{k=1}^{3} U^\nu_{ki} U^{\nu*}_{kj},
\end{align}

 Notice that in this model we consider the right handed neutrinos as singlets of $SU(2)$, so in the electroweak interaction basis there are no couplings between the right handed neutrinos $\nu_R$ with the $SU(2)$ gauge bosons $W$ and $Z$. Indeed, the right handed neutrinos only interact with the Higgs sector, i.e. with the Higgs boson and the Goldstone bosons.

The relation between the normalization of $F_L$ and ${\cal F}_L$ is:
\begin{equation}
F_L =  \frac{g^{3}}{32 \pi^{2} c_W}{\cal F}_L   \,.
\label{fnorm}
\end{equation}


\section{Appendix: MIA Form Factors (Feynman-'t Hooft gauge)} 
\label{app:FormFactors}
 The results of the MIA form factors to $\O(Y^2)$ in the Feynman-'t Hooft gauge are the following:
\begin{align}
F_{L}^{(2a)}&= \frac{1}{16 \pi^{2}} \frac{g m_{W}^{2}}{c_W} \left(Y^{}_{\nu} Y_{\nu}^{\dagger} \right)^{km} \big( -2D_{00}+p_{1}^{2}(D_0+D_1+D_{13}-D_{33})\big)(p_{2},0,p_{1},m_{W},0,M_{R},0) \,, \nonumber \\
F_{L}^{(2b)}&=\frac{1}{16 \pi^{2}} \frac{g m_{W}^{2}}{c_W} \left(Y^{}_{\nu} Y_{\nu}^{\dagger} \right)^{km} \left( -2D_{00}+p_{1}^{2}(D_0+D_1+D_{12}-D_{22})\right)(p_{2},p_{1},0,m_{W},0,M_{R},0) \,,
\nonumber \\
F_{L}^{(3a)}&= \frac{1}{16 \pi^{2}} 2gc_W m_{W}^{2} \left(Y^{}_{\nu} Y_{\nu}^{\dagger} \right)^{km} \big( C_{0}(p_2,p_1,M_R,m_W,m_W)  \,,\nonumber\\
& +(2D_{00}-p_{1}^{2}D_{2})(0,p_{2},p_{1},0,M_{R},m_{W},m_{W})   \big)\,,
\nonumber \\ 
 F_{L}^{(4a)}&= \frac{1}{16 \pi^{2}} g s_{W}^{2} m_W m_Z \left(Y^{}_{\nu} Y_{\nu}^{\dagger} \right)^{km} C_0(p_{2},p_{1},M_{R},m_W,m_W) \,,
\nonumber \\
F_{L}^{(5a)}&= \frac{1}{16 \pi^{2}} g s_{W}^{2} m_W m_Z \left(Y^{}_{\nu} Y_{\nu}^{\dagger} \right)^{km} C_0(p_{2},p_{1},M_{R},m_W,m_W) \,,
\nonumber \\
F_{L}^{(6a)}&=-\frac{1}{16 \pi^{2}} \frac{2g}{c_W} \left( -\frac{1}{2}+s_{W}^{2} \right) \left(Y^{}_{\nu} Y_{\nu}^{\dagger} \right)^{km} C_{00}(p_{2},p_{1},M_{R},m_W,m_W)\,,
 \nonumber \\
F_{L}^{(7a)}&=-\frac{1}{16 \pi^{2}} \frac{2g m_{W}^{2}}{c_W} \left( -\frac{1}{2}+s_{W}^{2} \right) \left(Y^{}_{\nu} Y_{\nu}^{\dagger} \right)^{km} \frac{m_{k}^{2}}{m_{k}^{2}-m_{m}^{2}} C_{2}(0,p_2,0,M_R,m_W)   \,,
\nonumber \\
F_{L}^{(8a)}&= -\frac{1}{16 \pi^{2}} \frac{g}{c_W} \left( -\frac{1}{2}+s_{W}^{2} \right) \left(Y^{}_{\nu} Y_{\nu}^{\dagger} \right)^{km} \frac{m_{k}^{2}}{m_{k}^{2}-m_{m}^{2}} B_1(p_2,M_{R},m_W) \,,
\nonumber \\
F_{L}^{(9a)}&= -\frac{1}{16 \pi^{2}} \frac{2g m_{W}^{2}}{c_W} \left( -\frac{1}{2}+s_{W}^{2} \right) \left(Y^{}_{\nu} Y_{\nu}^{\dagger} \right)^{km} \frac{-m_{m}^{2}}{m_{k}^{2}-m_{m}^{2}} C_{2}(0,p_3,0,M_R,m_W) \,,
\nonumber \\
F_{L}^{(10a)}&= -\frac{1}{16 \pi^{2}} \frac{g}{c_W} \left( -\frac{1}{2}+s_{W}^{2} \right) \left(Y^{}_{\nu} Y_{\nu}^{\dagger} \right)^{km} \frac{-m_{m}^{2}}{m_{k}^{2}-m_{m}^{2}} B_1(p_3,M_{R},m_W) \,.
\end{align}
All the remaining diagrams are of $\O(m_{lep}^{2})$, and since we are neglecting  the lepton masses in our computation they will provide vanishing contributions to the form factor. Specifically, these vanishing diagrams are:
\be
F_L^{(1a)}, F_L^{(1b)},  F_L^{(1c)},   F_L^{(1d)}, F_L^{(4b)}, F_L^{(5b)}, F_L^{(6b)},  F_L^{(6c)},   F_L^{(6d)},
F_L^{(8b)},   F_L^{(8c)},   F_L^{(8d)}, F_L^{(10b)},   F_L^{(10c)},   F_L^{(10d)}\,. \nonumber
\ee
The results of the MIA form factors to $\O(Y^4)$ in the Feynman-'t Hooft gauge are the following:
\begin{align}
F_L^{(1e)}&= \frac{1}{16 \pi^{2}} \frac{g}{2c_W} v^{2} \left(Y^{}_{\nu} Y_{\nu}^{\dagger} Y^{}_{\nu} Y_{\nu}^{\dagger} \right)^{km} C_{0}(p_{2},p_{1},m_W,M_{R},M_R)\,,
\nonumber \\
F_{L}^{(6e)}&= \frac{1}{16 \pi^{2}} \frac{g}{c_W} \left( 1-2s_{W}^{2} \right) v^{2} \left(Y^{}_{\nu} Y_{\nu}^{\dagger} Y^{}_{\nu} Y_{\nu}^{\dagger} \right)^{km} D_{00}(0,p_{2},p_{1},M_R,M_{R},m_W,m_W) \,,
\nonumber \\
F_{L}^{(8e)}&= -\frac{1}{16 \pi^{2}} \frac{g}{c_W} \left( -\frac{1}{2}+s_{W}^{2} \right) v^{2} \left(Y^{}_{\nu} Y_{\nu}^{\dagger} Y^{}_{\nu} Y_{\nu}^{\dagger} \right)^{km} \frac{m_{k}^{2}}{m_{k}^{2}-m_{m}^{2}} C_2(0,p_2,M_R,M_{R},m_W) \,,
\nonumber \\
F_{L}^{(10e)}&= -\frac{1}{16 \pi^{2}} \frac{g}{c_W} \left( -\frac{1}{2}+s_{W}^{2} \right) v^{2} \left(Y^{}_{\nu} Y_{\nu}^{\dagger} Y^{}_{\nu} Y_{\nu}^{\dagger} \right)^{km} \frac{-m_{m}^{2}}{m_{k}^{2}-m_{m}^{2}} C_2(0,p_3,M_R,M_{R},m_W) \,.
\end{align}
 
Regarding the divergences in the MIA computation in the Feynman-'t Hooft gauge we have found the following:
1)  to $\O(Y^2)$ 
the only divergent diagrams are (6a), (8a) and (10a), and we have checked that all these divergences cancel out when adding the three diagrams. So, our final result to $\O(Y^2)$ is finite. 2) to $\O(Y^4)$ all the loop functions are finite, and therefore all the diagrams are also finite. In summary, we have checked that the total form factor, $F_L^{\rm MIA}$, is finite for an arbitrary $p_1^2$.
 
 Finally, notice that these formulas are valid for the degenerate $M_{R_i}=M_R$ case. 
Nevertheless, they can be easily generalized to the non-degenerate case, as explained in~\cite{Arganda:2017vdb}. 
For example, it would be enough to change 
\begin{align}
(Y_\nu Y_\nu^\dagger)^{km} C_\alpha(p_2,p_1,M_R,m_W,m_W) 
&\rightarrow (Y_\nu^{ka} Y_\nu^{\dagger am}) C_\alpha(p_2,p_1,M_{R_a},m_W,m_W)\,, \nonumber\\
(Y_\nu Y_\nu^\dagger Y_\nu Y_\nu^\dagger)^{km} C_\alpha(p_2,p_1,m_W,M_R,M_R) 
&\rightarrow	(Y_\nu^{ka} Y_\nu^{\dagger ai} Y_\nu^{ib} Y_\nu^{\dagger bm}) C_\alpha(p_2,p_1,m_W,M_{R_a},M_{R_b})\,,
\end{align}
and similarly for all the other loop functions and terms. 


\section{Appendix: Large $\boldsymbol{M_R}$ expansion of the loop integrals}   
\label{Expansions}
Here we summarize the results of the large $M_R$ expansion for all the 
one-loop functions entering in the calculation of the on-shell effective vertex. Concretely, the ones involved in the Feynman-'t Hooft gauge that are given in Eq.~\eqref{Veffgeneric}. We use here the same notation as in the text, i.e, we use a  hat to denote the functions when evaluated at on-shell external $Z$ boson with $p_1^2=m_Z^2$. Besides, we neglect the lepton masses in all these one-loop functions, and provide their main result from the large $M_R$ expansion, namely, by keeping just the relevant terms that lead to the first order contribution in the effective vertex, i.e. the ${\cal O} (v^2/M_{R}^{2})$ term in Eq.~\eqref{generic}. We also use here the same notation as in text with  $\bf{M_R}$ in boldface to mean that we are performing the large $M_R$ expansion of the given function.

We find the following results: 
\begin{align}
& B_{1}\left({\bf M_R},m_W\right) = -\frac{\Delta }{2} -\frac{3}{4} +\frac{1}{2}\log \left(\frac{M_{R}^{2}}{\mu^{2}}\right) -\frac{m_{W}^{2}}{2 M_{R}^{2}} \left(2 \log \left(\frac{m_{W}^{2}}{M_{R}^{2}}\right)+1\right) \,, \nonumber\\
& {\hat C}_{0}\left(p_2,p_1,{\bf M_{R}},m_W,m_W\right) = \frac{1}{M_{R}^{2}} \left( 2(4r-1)^{\frac{1}{2}}  \arctan \Big[(4r-1)^{-\frac{1}{2}}\Big]-1+\log \left(\frac{m_{W}^{2}}{M_{R}^{2}}\right)  \right) \,, \nonumber\\
&{\hat C}_{0}\left(p_2,p_1,m_W,{\bf M_R},{\bf M_R}\right) = -\frac{1}{M_{R}^{2}}\,,  \nonumber\\
&{\hat C}_{00}\left(p_2,p_1,{\bf M_{R}},m_W,m_W\right) = \frac{\Delta }{4} +\frac{3}{8} -\frac{1}{4}\log \left(\frac{M_{R}^{2}}{\mu^{2}}\right)  +\frac{m_{W}^{2}}{72 M_{R}^{2}} \left( \left(6-r^{-1}\right)\left( 6\log \left(\frac{m_{W}^{2}}{M_{R}^{2}}\right)-5 \right) \right.  \nonumber\\
& \hspace{50mm} \left. +12 \left(4-r^{-1}\right)(4r-1)^{\frac{1}{2}}  \arctan \Big[(4r-1)^{-\frac{1}{2}}\Big] \right) \,, \nonumber\\
& C_{2}\left(0,{\bf M_{R}},m_W\right) = -\frac{1}{2 M_{R}^{2}} \left(1+\log \left(\frac{m_{W}^{2}}{M_{R}^{2}}\right) \right) \,, \nonumber\\
& C_{2}\left({\bf M_R},{\bf M_{R}},m_W\right) = \frac{1}{2 M_{R}^{2}} \,, \nonumber\\
&({\hat D}_{0}+{\hat D}_{1}+{\hat D}_{13}-{\hat D}_{33})\left(p_2,0,p_1,m_W,0,{\bf M_{R}},0\right) = 
 \nonumber\\
&\hspace{3mm}-\frac{1}{2m_Z^2 M_R^2} -\frac{1+r}{m_Z^2 M_R^2} \bigg\{(2 r+1) \text{Li}_2(r+1)-2 \left(\frac{\pi ^2}{6} (2 r+1)-1-i \pi \right)+\left(r+\frac{1}{2}\right) \log ^2(r) \nonumber\\
&\hspace{3mm}+ \big(2+i \pi  (2 r+1)\big) \log (r)\bigg\}   = ({\hat D}_{0}+{\hat D}_{1}+{\hat D}_{12}-{\hat D}_{22})\left(p_2,p_1,0,m_W,0,{\bf M_{R}},0\right)\,, \nonumber\\
& {\hat D}_{00}\left(p_2,0,p_1,m_W,0,{\bf M_{R}},0\right)= {\hat D}_{00}\left(p_2,p_1,0,m_W,0,{\bf M_{R}},0\right)=  \nonumber\\
& \hspace{3mm} -\frac{1}{24 M_R^2} \bigg\{-6 \log \left(\frac{m_W^2}{M_R^2}\right)+12 (r+1) r \text{Li}_2(r+1)-4 \pi ^2 r^2+6 \left(2 i \pi  r^2+2 i \pi  r+2 r+1\right) \log (r) \bigg\} \nonumber\\
& \hspace{3mm}-\frac{1}{24 M_R^2} \bigg\{-4 \pi ^2 r+12 i \pi  r+12 r+6 (r+1) r \log ^2(r)+6 i \pi +9 \bigg\}  \,,\nonumber\\
& {\hat D}_{00}\left(0,p_2,p_1,0,{\bf M_{R}},m_W,m_W\right) = \nonumber\\
&\hspace{.3cm}\frac{1}{M_{R}^{2}} \bigg\{ \frac{1}{4} \log \left(\frac{m_W^2}{M_R^2}\right)  -\frac{3}{8}  +\frac{r}{2} +\frac{1-2r}{2}(4r-1)^{\frac{1}{2}}  \arctan \Big[(4r-1)^{-\frac{1}{2}}\Big]   +r^{2}  \arctan^{2}\Big[ (4r-1)^{-\frac{1}{2}}\Big] \bigg\} \,, \nonumber\\
& {\hat D}_{2}\left(0,p_2,p_1,0,{\bf M_{R}},m_W,m_W\right) =\nonumber\\
&\hspace{.3cm} \frac{2}{m_{Z}^{2}M_{R}^{2}} \bigg\{ -4 r  \arctan^{2} \Big[(4r-1)^{-\frac{1}{2}}\Big] +2(4r-1)^{\frac{1}{2}}   \arctan \Big[(4r-1)^{-\frac{1}{2}}\Big]  -1 \bigg\}  \,,  \nonumber\\
& {\hat D}_{00}\left(0,p_2,p_1,{\bf M_R},{\bf M_{R}},m_W,m_W\right) = -\frac{1}{4 M_{R}^{2}}\,.
\end{align}
In these above formulas  we have denoted the $W$ and $Z$ squared mass ratio by $r=c_W^2=m_{W}^{2}/m_{Z}^{2}$ to shorten the result, and $m_W=gv/\sqrt{2}$. Besides, $\mu$ is the usual regularization scale of dimensional regularization, and the divergence  $\Delta$ is defined in Eq.~(\ref{DeltaDiv}).

Finally, for completeness, we also provide here the results for the additional loop functions in the zero external momenta case which are needed to get the results of the following appendix. These are:
\begin{align}
& C_{0}\left({\bf M_{R}},m_W,m_W\right) = \frac{1}{M_{R}^{2}} \left(1+\log \left(\frac{m_{W}^{2}}{M_{R}^{2}}\right) \right)  \,,\nonumber\\  
& C_{0}\left(m_W,{\bf M_R},{\bf M_R}\right) = -\frac{1}{M_{R}^{2}} \,,   \nonumber\\ 
& C_{00}\left({\bf M_{R}},m_W,m_W\right) = \frac{\Delta }{4} +\frac{3}{8} -\frac{1}{4}\log \left(\frac{M_{R}^{2}}{\mu^{2}}\right) +\frac{m_{W}^{2}}{4 M_{R}^{2}} \left(2 \log \left(\frac{m_{W}^{2}}{M_{R}^{2}}\right)+1\right) \,, \nonumber\\
& D_{00}\left(m_W,0,{\bf M_{R}},0\right) = \frac{1}{4 M_{R}^{2}} \log \left(\frac{m_{W}^{2}}{M_{R}^{2}}\right) \,,  \nonumber\\ 
& D_{00}\left(0,{\bf M_{R}},m_W,m_W\right) = \frac{1}{4 M_{R}^{2}} \left(1 +\log \left(\frac{m_{W}^{2}}{M_{R}^{2}}\right) \right) \,, \nonumber\\
& D_{00}\left({\bf M_R},{\bf M_{R}},m_W,m_W\right) = -\frac{1}{4 M_{R}^{2}}   \,.
\end{align}

 

\section{Appendix: The one-loop effective vertex $\boldsymbol{Z\ell_k\ell_m}$ at zero external momenta} 
\label{app:Zpenguins}
As we have mentioned in the text, it is an interesting exercise to evaluate the one-loop effective vertex $Z\ell_k\ell_m$ at zero external momenta, $V^{\rm eff}_{Z\ell_k\ell_m}|_{p_{\rm ext}^2=0}$ with $p_{\rm ext}^2=0$ meaning $p_1^2=p_2^2=p_3^2=0$, and find out if this is a valid and accurate result to be used inside a physical observable, like the LFV $Z$ partial decay width, or other low energy LFV processes that can be mediated by a $Z$ boson. In principle, one would naively expect that this effective vertex could provide a good approximation to the Z-penguin mediated  contributions in low energy observables, like LFV three body lepton decays, $\ell_m \to 3 \ell_k$, $\mu-e$ conversion in heavy nuclei and others. In those cases, working in the limit of very small transfer momentum at the intermediate $Z$ boson propagator is a good approximation, and therefore to provide a simple formula for the $V^{\rm eff}_{Z\ell_k\ell_m}|_{p_{\rm ext}^2=0}$ vertex seems to be useful. However, we have found that it is indeed not the case, since our results show explicitly that it is a gauge dependent quantity and cannot be used separately from the other contributions in these low energy observables, like the photon-penguin contributions, box diagrams and others. 

We present next our analytical results for this  $V^{\rm eff}_{Z\ell_k\ell_m}|_{p_{\rm ext}^2=0}$ vertex with several gauge choices. Firstly, we find that the ${\cal O}(Y_\nu^4)$ contribution is gauge independent and coincides in all covariant gauges with the result obtained for the on-shell case, i.e, we get the same analytical result as in Eq.~\eqref{veffsimple}. Secondly, we find that the result of the ${\cal O}(Y_\nu^2)$ contribution in the unitary gauge is divergent. Specifically, the divergence is:
\ba
 F_L^{{\rm UG}(\Delta)}|_{p_{\rm ext}^2=0} 
&=& \frac{g}{16 \pi^2 c_W} \frac{\Delta}{4} 
 \left(Y^{}_{\nu} Y_{\nu}^{\dagger} \right)^{km}.
\label{UGdiv}
\ea
This divergent result, shows that the unitary gauge does not provide a physical result for this vertex when the external momenta are set to zero. Furthermore, we have also studied in detail the particular case of $\ell_m \to 3 \ell_k$ decays and we have checked by an explicit computation that by adding all the contributions, this divergence of the $Z$-penguin in Eq.~\eqref{UGdiv} cancels with the divergent contributions from the photon penguin and box diagrams, providing a finite result for the partial width of $\ell_m \to 3 \ell_k$ decays. However, this zero external momenta result of the UG cannot be used for the $Z$ decays.

Thirdly, for the case of the Feynman-'t Hooft gauge (FH), we get the following finite result:
\be
V^{\rm eff}_{Z\ell_k\ell_m}|_{p_{\rm ext}^2=0}^{\rm FH}  
= \frac{g}{16 \pi^2 c_W} \left[
 \frac{m_W^2}{M_R^2} \left( \frac{5}{2} + \frac{3}{2} \log \left( \frac{m_W^2}{M_R^2} \right) \right)
 \left(Y^{}_{\nu} Y_{\nu}^{\dagger} \right)^{km}
 - \frac{v^2}{2M_R^2} 
\left( Y^{}_{\nu} Y_{\nu}^{\dagger}Y^{}_{\nu} Y_{\nu}^{\dagger} \right)^{km}  
\right]  \,.
\label{veffFH}
\ee   
We have checked that this result is in agreement with the $F_Z^{ll'}$ of~\cite{Ilakovac:1994kj}, given in the physical neutrino mass basis, once we take  the heavy neutrino limit with $m_N \gg v$,  and after using the following relations,
\begin{align}
\sum_{i\in{\rm Heavy}} B_{\ell_kn_i}B^*_{\ell_mn_i}&\simeq \frac{v^2}{m_N^2} \big(Y_\nu^{} Y_\nu^\dagger\big)^{km} \,,\\
\sum_{i,j\in{\rm Heavy}} B_{\ell_kn_i}C_{n_in_j}B^*_{\ell_mn_j}&\simeq \frac{v^4}{m_N^4} \big(Y_\nu^{} Y_\nu^\dagger Y_\nu^{} Y_\nu^\dagger\big)^{km} \,.
\end{align}

  We have also checked that the logarithmic contributions within Eq.~\eqref{veffFH} coming from the specific one-loop diagrams with only one neutrino propagator in the loops, i.e., from diagrams of type (3) through type (10), are in agreement  
with the logarithmic contribution provided in Eqs.~(11-14) of~\cite{Abada:2015zea} in the proper limit of heavy neutrinos, i.e. for $x_i=m_{n_i}^2/ m_W^2 \gg 1$. Notice that this reference~\cite{Abada:2015zea} presents their results as a Taylor expansion around zero external $Z$ momentum, therefore they cannot be compared with our on-shell $Z$ results, but just with our zero external $Z$ momentum results.  
Concretely, we find for this partial subset of diagrams of the Feynman-'t Hooft gauge the following result for the logarithmic term to 
${\cal O}(Y_{\nu}Y_{\nu}^{\dagger})$:
\begin{equation}
V^{\rm eff}_{Z\ell_k\ell_m}|_{p_{\rm ext}^2=0}^{\rm FH} 
({\rm only \,\, diags \,\, 3's+..+10's}) 
= \frac{g}{16 \pi^2 c_W} ~
 \frac{m_W^2}{M_R^2} ~\frac{5}{2} \log \left( \frac{m_W^2}{M_R^2} \right) 
 \left(Y^{}_{\nu} Y_{\nu}^{\dagger} \right)^{km} +\dots
   \,,
\label{veffFH-partial}
\end{equation}   
which is in agreement with the $\log x_i$ contribution in Eqs.~(11-14) of~\cite{Abada:2015zea}. 
We cannot perform a complete comparison with~\cite{Abada:2015zea}, i.e. including other contributions like the finite non-logarithmic term nor the contributions from the full set of diagrams containing two neutrino propagators (heavy-heavy, light-heavy, light-light), since they are not provided in this reference~\cite{Abada:2015zea} in the needed limit of the heavy $n_i$, namely, for $x_i=m_{n_i}^2/ m_W^2 \gg 1$.   

And, finally, for the case of an arbitrary $R_\xi$ gauge we get the following finite result:
\be
\hspace{-.2cm}V^{\rm eff}_{Z\ell_k\ell_m}|_{p_{\rm ext}^2=0}^{R_\xi}  
= \frac{g}{16 \pi^2 c_W} \left[
 \frac{m_W^2}{M_R^2} \left( h(\xi)  + \frac{3}{2} \log \left( \frac{m_W^2}{M_R^2} \right) \right)
 \left(Y^{}_{\nu} Y_{\nu}^{\dagger} \right)^{km}
 - \frac{v^2}{2M_R^2} 
\left( Y^{}_{\nu} Y_{\nu}^{\dagger}Y^{}_{\nu} Y_{\nu}^{\dagger} \right)^{km}  
\right]  \,,
\label{veffxi}
\ee  
where the $\xi$ parameter dependence is included in the function:
\be 
 h(\xi)= \frac{\xi+3}{4}+\frac{3}{2} \frac{\xi \log \xi}{\xi-1}\,.
\ee
 Notice that we expanded the one-loop functions in the large $M_R$ limit taking into account the gauge-fixing parameter: $M_R \gg m_W, \sqrt{\xi}m_W$.

The previous result of Eq.~(\ref{veffxi}) clearly demonstrates that the one-loop effective vertex at zero external momenta is not a physical quantity since it is manifestly gauge dependent. In this $R_\xi$ gauge case, we have also checked by an explicit computation of all the contributions to the $\ell_m \to 3 \ell_k$ decays that the previous  $\xi$
dependence from the $Z$ penguin is cancelled by the photon penguin and boxes contributions, leading to a gauge invariant result, as it must be. 
On the other hand, we would like to emphasize that,   although the results in Eqs.~\eqref{veffFH} and \eqref{veffxi} can be useful for a discussion of the (gauge dependent) $Z$ penguin contribution in a low energy processes, they cannot be used for the $Z$ decays case, since as  proven here, they are gauge dependent.


\end{appendix}


\bibliography{bibliography}

\end{document}